\documentclass[12pt,graphicx,subfigure,axodraw]{article}
\usepackage{amsfonts}
\usepackage{amssymb}
\usepackage{epsfig}
\usepackage{psfig}
\usepackage{psfrag}
\usepackage{dsfont}

\usepackage{axodraw}

\newcommand{\beqn}{\begin{eqnarray}}
\newcommand{\eeqn}{\end{eqnarray}}
\newcommand{\be}{\begin{equation}}
\newcommand{\ee}{\end{equation}}
\newcommand{\non}{\nonumber \\}

\newcommand{\te}{\theta}
\newcommand{\thb}{\bar\theta}

\newcommand{\ca}{{\cal A}}

\newcommand{\cl}{{\cal L}}
\newcommand{\cf}{{\cal F}}

\newcommand{\co}{{\cal O}}

\def\st{Stueckelberg~}
\def\s1{$s_{\alpha}$}
\def\s2{$s_{\gamma}$}
\def\s3{$s_{\delta}$}
\def\c1{$c_{\alpha}$}
\def\c2{$c_{\gamma}$}
\def\c3{$c_{\delta}$}

\def\za{s^2\Gamma_Z^2 M_Z^{-2}}
\def\zb{s^2\Gamma_{{\rm Z}'}^2 M_{{\rm Z}'}^{-2}}

\begin{document}
\baselineskip 18pt

\thispagestyle{empty}

\begin{flushright}
\vspace{-3cm}
{\small DESY 05-044 \\
        MIT-CTP-3612 \\
        NUB-TH-3254 \\[-.15cm]
        hep-ph/0503208
}
\end{flushright}
\vspace{0cm}

\begin{center}
{\Large\bf
 Aspects of the Stueckelberg Extension }

\vspace{1cm}

{\bf Boris K\"ors}\footnote{e-mail: kors@lns.mit.edu}$^{,*}$
{\bf and Pran Nath}\footnote{e-mail: nath@neu.edu}$^{,\dag}$
\vspace{.5cm}

{\it
$^{\dag}$Department of Physics, Northeastern University \\
Boston, Massachusetts 02115, USA \\

$^*$II. Institut f\"ur Theoretische Physik, Universit\"at Hamburg\\
Luruper Chaussee 149,
22761 Hamburg, Germany\\

$^*$Center for Theoretical Physics,
Laboratory for Nuclear Science \\
and Department of Physics,
Massachusetts Institute of Technology \\
Cambridge, Massachusetts 02139, USA \\
}

\end{center}

\vspace{0.2cm}

\begin{center}
{\bf Abstract} \\
\end{center}
\vspace{-.2cm}
A detailed analysis of a Stueckelberg extension of the
electro-weak gauge group with an extra abelian
$U(1)_X$ factor is presented for the Standard Model as well as for the MSSM.
The extra gauge boson gets massive through a Stueckelberg type coupling to
a pseudo-scalar, instead of a Higgs effect.
This new massive neutral gauge boson Z$'$ has vector
and axial vector couplings uniquely different
from those of conventional extra abelian gauge bosons, such as appear e.g.\ in GUT models.
The extended MSSM furthermore contains two extra neutralinos and one extra neutral CP-even scalar,
the latter with a mass larger than that of the Z$'$.
One interesting scenario that emerges is an LSP that
is dominantly composed out of the new neutralinos, leading to a possible
new superweak candidate for dark matter.
We investigate signatures of the Stueckelberg extension
at a linear collider and discuss techniques for the detection of the expected
sharp Z$'$ resonance. It turns out that
the substantially modified forward-backward asymmetry around the Z$'$ pole
provides an important signal.
Furthermore, we also elaborate on generalizations of the minimal \st extension
to an arbitrary number of extra $U(1)$ gauge factors.

\clearpage \setcounter{footnote}{0}


\tableofcontents

\section{Introduction}

The Stueckelberg Lagrangian \cite{stueck} is a gauge invariant
kinetic term for a massive abelian vector field, that utilizes a
non-linear representation of the gauge transformation. The mass
term is made gauge invariant by coupling a massless gauge boson to
a real pseudo-scalar, which then transforms non-linearly, and in
unitary gauge is absorbed as the longitudinal mode of the massive
vector. As we shall point out below, gauge boson masses through
Stueckelberg couplings are ubiquitous in compactifications of
higher-dimensional string theory, supergravity, or even pure gauge
theory. From a model building perspective, the relevance of the
Stueckelberg mechanism lies in the fact that it provides an
opportunity alternative to the Higgs mechanism \cite{higgs} to
achieve gauge symmetry breaking without spoiling renormalizability
\cite{AML}. Since the minimal version of the Stueckelberg
mechanism only needs a single real scalar, which is absorbed by
the gauge boson with no other degrees of freedom left, it is
already clear that the Stueckelberg and the Higgs mechanism are
physically distinct. The main purpose of this paper is to discuss the
 most simple extensions of the electro-weak sector of the Standard
Model (SM) \cite{gws}, and its supersymmetric generalizations (SSM
or MSSM). The work presented here is a more detailed exposition
and extension of of two previous publications, where the \st
extension was first achieved \cite{kn1,kn2,Kors:2004iz}. In
particular, an analysis of the possibility of observation of \st
phenomena at linear colliders is also given.
\\


\subsection{The \st Lagrangian}

The prototype Stueckelberg Lagrangian couples one abelian vector
boson $A_\mu$ to one pseudo-scalar $\sigma$ in the following
way,\footnote{As is well known, the Stueckelberg mechanism can
actually be recovered in a rather singular limit of the Higgs
mechanism \cite{AML}, and it is useful to keep the comparison in
mind as we discuss the models based on the \st mechanism. This
similarity, however, is not so easily realized in the case of the
supersymmetric \st extension of the MSSM.}
\beqn
{\cal L} = -\frac14 {\cal F}_{\mu\nu}{\cal F}^{\mu\nu} - \frac12
(m A_\mu + \partial_\mu \sigma)(m A^\mu + \partial^\mu \sigma)\ .
\eeqn
It is gauge invariant if $\sigma$ transforms together with $A_\mu$
according to
\beqn
\delta A_\mu = \partial_\mu \epsilon \ , \quad
\delta \sigma = - m \epsilon \ .
\eeqn
Fixing the gauge by adding
\beqn
\cl_{\rm gf} = - \frac{1}{2\xi} \left( \partial_\mu A^\mu + \xi m \sigma \right)^2 \ ,
\eeqn
the total Lagrangian reads
\beqn
\cl+\cl_{\rm int}+\cl_{\rm gf} &=&
-\frac14 {\cal F}_{\mu\nu}{\cal F}^{\mu\nu} - \frac{m^2}{2} A_\mu A^\mu - \frac{1}{2\xi} (\partial_\mu A^\mu)^2
\non
&&
- \frac12 \partial_\mu \sigma \partial^\mu \sigma - \xi \frac{m^2}{2} \sigma^2
\eeqn
where the two fields have been decoupled, and renormalizability
and unitarity are manifest. To add interactions with fermions, one
may couple the vector field to a conserved current, adding the
interaction
\beqn
\cl_{\rm int} = g A_\mu J^\mu
\eeqn
with $\partial_\mu J^\mu = 0$.
\\


Let us mention here that regarding the extension of this mechanism
to non-abelian gauge theories, according to \cite{nonabelian}, a
non-abelian extension of the Stueckelberg Lagrangian leads to
violation of unitarity already at the tree-level, because the
longitudinal components of the vector fields cannot be decoupled
from the physical Hilbert space. The renormalizability of the
theory is then spoiled as well. Therefore, the Higgs of the SM
cannot be replaced by a Stueckelberg type of symmetry breaking.
Instead we will consider extensions of the SM or the MSSM which
involve extra $U(1)$ gauge factors beyond the $SU(3)_C\times
SU(2)_L\times U(1)_Y$ gauge symmetry of the SM, which will then be
assumed to couple to pseudo-scalars in the way of Stueckelberg.


\subsection{Stueckelberg in string theory and compactification}

One immediate way to see that Stueckelberg couplings appear in
dimensional reduction of supergravity from higher dimensions, and
in particular string theory, is to consider the reduction of the
ten-dimensional $N=1$ supergravity coupled to supersymmetric
Yang-Mills gauge fields \cite{Chamseddine:ez}, in the presence of
internal gauge fluxes. The ten-dimensional kinetic term for the
anti-symmetric 2-tensor $B_{IJ}$ involves a coupling to the
Yang-Mills Chern-Simons form, schematically $\partial_{[I} B_{JK]}
+ A_{[I} F_{JK]} +\frac23 A_{[I} A_J A_{K]}$, in proper units.
Dimensional reduction with a vacuum expectation value for the
internal gauge field strength, $\langle F_{ij} \rangle \not=0$,
leads to
\beqn
\partial_{\mu} B_{ij} + A_{\mu} F_{ij} ~\sim~ \partial_\mu \sigma + m A_\mu  \ ,
\eeqn
after identifying the internal components $B_{ij}$ with the scalar
$\sigma$ and the value of the gauge field strength with the mass
parameter $m$. Thus, $A_\mu$ and $\sigma$ have a Stueckelberg
coupling of the form $A_\mu \partial^\mu \sigma$. These couplings
play an important role in the Green-Schwarz anomaly cancellation
mechanism. In a four-dimensional theory abelian gauge symmetries
can have a triangle ABJ anomaly, if tr$\, Q\neq 0$ or tr$\,
Q^3\neq 0$. In a consistent string compactification, this ABJ
anomaly is cancelled by Green-Schwarz type contributions involving
the two terms $m A^\mu \partial_\mu \sigma + c\, \sigma F_{\mu\nu}
\tilde F^{\mu\nu}$ in the Lagrangian and the anomalous 3-point
function is proportional to the product of the two couplings, $m
\cdot c$, while the mass parameter in the Stueckelberg coupling is
only $m$. Therefore, any anomalous $U(1)$ will always get massive
through the Stueckelberg mechanism, since $m \cdot c\not=0$, but a
non-anomalous $U(1)$ can do so as well, if $m \not=0,\, c=0.$
Since we do not want to deal with anomalous gauge symmetries here,
we shall always assume that $m \not=0,\, c=0.$ The mass scale that
determines $m$ within models that derive from string theory can,
at leading order, also be derived from dimensional reduction. It
turns out to be proportional to the string or compactification
scale in many cases \cite{ghiletal}, but can in principle also be
independent \cite{Ibanez:1998qp}.
\\

The fact that an abelian gauge symmetry, anomalous or
non-anomalous, may decouple from the low energy theory via
Stueckelberg couplings was actually of great importance in the
construction of D-brane models with gauge group and spectrum close
to that of the SM \cite{Ibanez:2001nd}. Roughly speaking, these
D-brane constructions start with a number of unitary gauge group
factors $U(N)$, which are then usually broken to their subgroups
$SU(N)$ via Stueckelberg couplings,
\beqn
U(3) \times U(2) \times U(1)^2
~~\stackrel{\rm Stueckelberg}{\longrightarrow}~~ SU(3)_C\times SU(2)_L \times U(1)_Y\ .
\eeqn
The mass matrix for the abelian gauge bosons is then
block-diagonal, and only the SM survives. In order to ensure this
pattern, one has to impose a condition on the Stueckelberg
mass parameters, namely that the hyper charge
gauge boson does not couple to any axionic scalar and remains
massless \cite{Ibanez:2001nd}. In the language of these D-brane
models, we will here relax this extra condition, and explore the consequences
of letting the hyper charge gauge boson
mix with other abelian gauge factors beyond the SM gauge
group, which seems a very natural extension of the SM in this
frame work.
\\

In a much simpler framework, in the dimensional compactification
of abelian gauge theory on a circle, one can also demonstrate that
the higher Kaluza-Klein excitations of the vector field gain their
mass through a \st mechanism. For this purpose we consider a
five-dimensional abelian gauge field $A_I$, using coordinates $x_I
= (x_{\mu},y)$. The gauge kinetic energy including a gauge fixing
term is
\beqn
\cl_{5d} = -\frac{1}{4} \cf_{IJ}(z) \cf^{IJ} - \frac{1}{2\xi} (\partial_I A^I )^2\ .
\eeqn
We compactify the fifth dimension on a half circle ${\cal S}^1$ of
radius $R$ and expand the five-dimensional gauge field $A_I(x_I)
=(A_{\mu}(x_I), \sigma(x_I))$ in harmonics on the compactified
dimension,
\beqn
A_{\mu}(x_I) = \sum_{n=0}^{\infty} A_{\mu}^{(n)}(x_\mu) \xi_n(y)\ ,\quad
\sigma(x_I) =\sum_{n=0}^{\infty} \sigma^{(n)}(x_\mu) \eta_n(y)\ ,
\eeqn
where $\xi_n(y)$ and $\eta_n(y)$ are harmonic functions on the
interval $(0,2\pi R)$ with appropriate periodicity conditions. The
effective  Lagrangian in four dimensions is obtained  by
integration over the fifth dimension,
\beqn
\cl_{4d} &=& \sum_{n=0}^{\infty} \Bigg[ -\frac{1}{4} \cf_{\mu\nu}^{(n)} \cf^{\mu\nu(n)}
- \frac{1}{2} n^2 ( M A_{\mu}^{(n)} + n \partial_\mu \sigma^{(n)} )^2
\non
&&
-\frac{1}{2\xi} \Big[ (\partial_{\mu} A^{(n)\mu} )^2
                  +2n M \partial_\mu A^{(n)\mu} \sigma^{(n)} + M^2 (\sigma^{(n)})^2 \Big] \Bigg]\ ,
\label{a4}
\eeqn
where $M = 1/R$ is the inverse of the compactification radius. The
Stueckelberg mechanism is now manifest in the first
line of Eq.(\ref{a4}). Choosing the gauge $\xi =1$ one finds
that the bilinear terms involving $A_{\mu}^{(n)}$ and
$\sigma^{(n)}$ form a total divergence which can be discarded, and
the scalar fields $\sigma^{(n)}$ decouple from the vector fields.
One is thus left with one massless vector field and an infinite
tower of massive vector fields all of which gain masses by the
Stueckelberg mechanism. There is no Higgs phenomenon involved in
the generation of their masses.


\subsection{Overview and summary}

The rest of the paper is devoted to further development of the
Stueckelberg extension of the SM and of the MSSM, and applications
to a number of phenomena which have the possibility of being
tested in current and future experiment. The outline is as
follows: In section 2 we give a detailed discussion of the
extension of the SM electro-weak gauge group $SU(2)_L\times
U(1)_Y$ to $SU(2)_L\times U(1)_Y\times U(1)_X$. We show that the
Stueckelberg extension allows one to retain a massless mode which
is identified with the photon, while the remaining two vector
bosons become massive and correspond to the gauge bosons Z and
${\rm Z}'$. Several useful results relating the mass parameters
and the mixing angles are deduced and general formulae for the
neutral current couplings to fermions are deduced. In section 3 we
give a full analysis of the extension of MSSM to include a \st
$U(1)_X$ gauge group, where in addition to a gauge vector
multiplet for the $U(1)_X$ one also has a chiral multiplet that
involves the \st pseudo-scalar. The Stueckelberg extension here
reproduces the vector boson sector of the Stueckelberg extension
of the SM and in addition contains new states and interactions
including an additional spin zero state, an extra neutral gaugino
and an extra neutral chiral fermion. In this section we also
discuss the implications of including Fayet-Illiopoulos D-terms in
the analysis.
\\

In section 4 we discuss the implications and predictions of the
Stueckelberg extensions. We work out in detail the deviations from
the SM couplings in the neutral current sector and estimate the
size of the parameters in the mixing of the \st sector with the
SM. Consistency with current data translates into bounds on these
parameters. However, refined experiments should be able to discern
deviations from the SM, such as the presence of a sharp ${\rm Z}'$
resonance. A careful scanning of data will be needed to discern
such a resonance. An explicit analysis of the modifications of the
Z boson couplings and of the couplings of the ${\rm Z}'$ boson to
SM fermions shows that the ${\rm Z}'$ has decay signatures which
are very distinct from the Z, and the observation of such
signatures should uniquely identify the ${\rm Z}'$ boson. In this
section we also discuss the mixing of the CP-even spin zero state
from the Stueckelberg chiral multiplet with the two CP-even Higgs
of the SM producing a $3\times 3$ CP-even Higgs mass matrix.  In
the neutralino sector there are now two more neutral states
arising from the \st sector, which mix with the four neutralino
states from the  MSSM producing a $6\times 6$ neutralino mass
matrix. There exists a region of the parameter space where one of
the \st fermions is the LSP. This would have a drastic influence
on collider signals for supersymmetry. Similarly, the dark matter
relic abundance will be affected. Additional topics discussed in
section 4 include the decay of ${\rm Z}'$ into the hidden sector
fermions which tend to give it a significantly larger decay width
than what is allowed by the decays into the visible sector, and
the modification of the correction to $g_{\mu}-2$ by inclusion of
the ${\rm Z}'$ boson exchange.
\\

In section 5 a detailed investigation for testing the Stueckelberg
scenario at a linear collider is performed. We analyze the
$e^+e^-$ cross section into leptons and quarks and also the
forward-backward asymmetry $A_{fb}$. It is shown that in the
vicinity of the ${\rm Z}'$ resonance it deviates significantly
from the SM prediction and hence will be a good indicator for
discerning such a resonance. In section 6 we discuss briefly the
technique that may be used for the detection of an expected sharp
resonance for the Z$'$. In section 7 we include a generalization
of the minimal Stueckelberg extension with just one extra $U(1)$
to an arbitrary number of extra abelian factors. Section 8 is
devoted to conclusions.


\section{Stueckelberg extension of the Standard Model}

We now turn to the main subject of this paper, the minimal
extensions of the SM and the MSSM which involve Stueckelberg type
couplings, and their experimental signatures. We start naturally
with the SM, then discuss the supersymmetrized version for the
MSSM, and afterward discuss the observable consequences. In any
case, since the Stueckelberg is only compatible with abelian gauge
symmetries, the minimal model that carries non-trivial structure
is obtained by adding an abelian gauge group factor $U(1)_X$ to
the SM gauge group, extending it to $SU(3)_C\times SU(2)_L\times
U(1)_Y\times U(1)_X$.\footnote{In principle, one could also just
consider a Stueckelberg coupling for only the hyper charge gauge
boson, but this would ultimately give a non-vanishing mass to the
photon, which is unacceptable.}
Then, all the abelian factors, i.e. hyper charge and $U(1)_X$, can
couple to a real pseudo-scalar $\sigma$ in the way of the
Stueckelberg mechanism.\footnote{We frequently call this scalar an
axionic scalar because of its pseudo-scalar nature, which does not
imply that it couples to QCD gauge fields in the way of the usual
QCD axion.}
We call this model the StSM (or StMSSM for the supersymmetric
version). In greater generality, one can of course add any number
of abelian factors to the SM, and have all the abelian gauge
bosons couple to any number of pseudo-scalars. In many string
theoretic models based on D-branes and orientifolds, there is
indeed a number of such gauge factors and scalars present, the
maximum multiplicity being restricted by topological properties of
the compactification space. We will come back to this option in
section \ref{sec7}.
\\

To start with the StSM \cite{kn2}, let $A_{\mu}^a$, $a=1,2,3$, be the
vector fields in the adjoint of $SU(2)_L$, with field strength
$F^a_{\mu\nu}$, $B_{\mu}$ the hyper charge vector with field
strength $B_{\mu\nu}$, and $\Phi$ be the Higgs
doublet.\footnote{The color $SU(3)_C$ factor of the gauge group
will be irrelevant for most of what we have to say.}
Then the relevant part of the SM Lagrangian is given by
\beqn \label{sm}
{\cal L}_{\rm SM}= -\frac{1}{4} {\rm tr}\, F_{\mu\nu}F^{\mu\nu} -\frac{1}{4} B_{\mu\nu}B^{\mu\nu}
+ g_2 A^a_\mu J^{a\mu}_2 + g_Y B_\mu J^\mu_Y
- D_{\mu}\Phi^{\dagger} D^{\mu}\Phi - V(\Phi^{\dagger}\Phi)\ ,
\eeqn
where $D_{\mu}\Phi$  is the gauge covariant derivative. For the
minimal Stueckelberg extension of this Lagrangian, we add the
degrees of freedom of one more abelian vector field $C_\mu$ for
the $U(1)_X$, with field strength $C_{\mu\nu}$, and one
pseudo-scalar $\sigma$.
For the scalar field $\sigma$ we assume that it will
have Stueckelberg couplings to all the abelian gauge bosons,
$B_\mu$ and $C_\mu$. For the Higgs scalar $\Phi$ we assume that it is
neutral under the $U(1)_X$, which just means that $C_\mu$ does not
appear in $D_\mu \Phi$. This is an assumption somehow
``orthogonal'' to the starting point of most models with so-called
$U(1)'$ gauge symmetries beyond the SM \cite{reviews}. 
There, the gauge symmetry
of the extra factor is broken by an extended Higgs effect. 
In our model, all non-trivial modification of the SM results from the
Stueckelberg coupling and is mediated by the axion $\sigma$. Thus,
the Lagrangian of Eq.(\ref{sm}) is extended to the StSM by
\beqn
\cl_{\rm StSM} = {\cal L}_{\rm SM}+{\cal L}_{\rm St}
\eeqn
with
\beqn
{\cal L}_{\rm St} = -\frac{1}{4} C_{\mu\nu}C^{\mu\nu}
+ g_X C_\mu J^\mu_X
- \frac{1}{2} (\partial_{\mu}\sigma  + M_1 C_{\mu} + M_2 B_{\mu})^2\ .
\eeqn
Up to this point, we have not specified the charges of the SM
fermions, which in principle could carry charges under $U(1)_X$.
Later, we will, however, abandon this possibility. Furthermore,
there may also be a sector that is hidden with respect to the SM
gauge symmetries, i.e. neutral with respect to $SU(2)_L\times
U(1)_Y$, but charged under $U(1)_X$, and thus enters $J_X^\mu$. In
such a case, the Stueckelberg coupling would be the only way to
communicate to the hidden sector. The extended non-linear gauge
invariance now reads
\beqn
\delta_Y B_{\mu} = \partial_{\mu}\lambda_Y\ , \quad
\delta_Y \sigma = - M_2 \lambda_Y\ ,
\eeqn
for the hyper charge, and
\beqn
\delta_X C_{\mu} = \partial_{\mu}\lambda_X\ ,\quad
\delta_X \sigma = - M_1 \lambda_X\ .
\eeqn
for $U(1)_X$. To decouple the two abelian gauge bosons from
$\sigma$, one has to add a similar gauge fixing term as in the
previous section with only one vector field. Furthermore, one has
to add the standard gauge fixing terms for the charged gauge
bosons to decouple from the Higgs.


\subsection{Masses for the neutral vector bosons}

After spontaneous electro-weak symmetry breaking the mass terms,
with mass-squared matrix $M^{[1]2}_{ab}$ (upper index $[1]$ for
spin 1) for the neutral vector bosons $V_{\mu a} = ( C_{\mu},
B_{\mu}, A_{\mu}^3 )_a$, take the form
\beqn
- \frac12 \sum_{a,b=1}^3 V_{\mu a} M^{[1]2}_{ab} V^\mu_b \ ,
\eeqn
where
\beqn \label{vecmass}
M^{[1]2}_{ab} =
\left[\matrix{ M_1^2  &  M_1M_2  &  0\cr
M_1M_2 & M_2^2 + \frac14 g_Y^2 v^2 & - \frac14 g_Yg_2 v^2  \cr
0 & -\frac14 g_Yg_2 v^2 & \frac14 g_2^2 v^2  }\right]
\eeqn
where $g_2$ and $g_Y$ are the $SU(2)_L$ and $U(1)_Y$ gauge
coupling constants, and are normalized so that $M_W^2=g_2^2v^2/4$.
From det$(M^{[1]2}_{ab})=0$ it is easily seen that one eigenvalue
is zero, whose eigenvector we identify with the photon
$A_\mu^\gamma$,\footnote{Note that we succeeded in obtaining a
massless photon, while previous attempts to obtain a Stueckelberg
extension of the SM failed in this respect \cite{Kuzmin:pg}. The
basic reason for the difficulty in keeping the photon massless
arose because there was an extra axion field which allowed the
photon to generate a tiny mass. In our analysis we have an extra
axion field $\sigma$ and two gauge bosons, $B_{\mu}$ and
$C_{\mu}$. Thus, after absorption of the axion we are indeed left
with a strictly massless photon field.}
and  the remaining two eigenvalues are the roots
\beqn
M^2_{\pm} &=&
\frac{1}{2} \Bigg[ M_1^2 + M_2^2 +\frac14 g_Y^2 v^2 + \frac14 g_2^2 v^2
\\
&& \hspace{.5cm}
\pm \Big[ \big( M_1^2 + M_2^2 +\frac14 g_Y^2 v^2 + \frac14 g_2^2 v^2 \big)^2
-\left( M_1^2(g_Y^2+g_2^2)v^2 +g_2^2 M_2^2 v^2 \right) \Big]^{\frac12} ~~ \Bigg]
\label{zmasses}
\nonumber
\eeqn
Obviously,
\beqn
M^2_+ = M_1^2+M_2^2 + {\cal O} (v^2)\ , \quad
M^2_- = {\cal O} (v^2)\ .
\eeqn
We, therefore, identify the mass eigenstate with mass squared
$M_-^2$ with the Z-boson, and call the mass eigenstate with
eigenvalue $M_+^2$ the Z$'$-boson. The diagonal matrix of
eigenvalues $E^{[1]}$ and the three eigenstates $E^{[1]}_\mu$ are
denoted
\beqn
E^{[1]} = {\rm diag} (M^2_{{\rm Z}'} , M^2_{\rm Z} , 0)= {\rm diag} (M^2_+ , M^2_- , 0)\ , \quad
E^{[1]}_\mu = ( {\rm Z}'_\mu , {\rm Z}_\mu , A^\gamma_\mu )^T\ .
\eeqn
Thus, we have
\beqn
V_{\mu a} = \sum_{b=1}^3 {\cal O}^{[1]}_{ab} E^{[1]}_{\mu b} \ ,\quad
\sum_{b,c=1}^3 {\cal O}^{[1]}_{ba} M^{[1]2}_{bc} {\cal O}^{[1]}_{cd} = E^{[1]}_{ad} \ , \quad
\label{diag}
\eeqn
for some orthogonal transformation matrix ${\cal O}^{[1]}_{ab}$.
One can actually solve for it explicitly. We use the parametrization
\beqn 
{\cal O}^{[1]}=
\left[\matrix{ \cos\psi \cos\phi -\sin\theta\sin\phi\sin\psi &
-\sin\psi \cos\phi -\sin\theta\sin\phi\cos\psi & -\cos\theta \sin\phi\cr
\cos\psi \sin\phi +\sin\theta\cos\phi\sin\psi &
-\sin\psi \sin\phi +\sin\theta\cos\phi\cos\psi & \cos\theta \cos\phi\cr
-\cos\theta\sin\psi & -\cos\theta\cos\psi & \sin\theta }\right]
\nonumber
\eeqn
Inverting the relation, one finds immediately
\beqn \label{taninv}
\tan (\phi) &=& \frac{M_2}{M_1} ~=~ \delta \ , \quad
\tan (\theta) ~=~ \frac{g_Y}{g_2}\cos(\phi) ~=~ \tan(\theta_W)\cos(\phi) \ .
\eeqn
Expressing $\tan(\psi)$ is a bit more tedious, and we use
\beqn \label{tanpsi}
\tan (\psi) &=& \frac{\tan(\theta)\tan(\phi)M_{{\rm W}}^2}
                     {\cos(\theta)(M_{{\rm Z}'}^2-M_{\rm W}^2(1+\tan^2(\theta)))}
\ ,
\eeqn
where $M_{\rm W}=g_2 v/2,\ \tan(\theta_W)=g_Y/g_2$. One can define
the two independent parameters to describe the StSM extension,
\beqn
\delta = \frac{M_2}{M_1}\ , \quad M^2 = M_1^2 + M_2^2 \ .
\eeqn
Effectively, $M$ is the overall mass scale of the new physics, and
$\delta$ the parameter that measures the strength of its coupling
to the SM. In the limit $\delta \rightarrow 0$, where the SM and a
decoupled abelian vector boson $C_\mu$ with mass $M$ are
recovered, one has
\beqn
\tan (\phi)\, , \ \tan (\psi) ~\longrightarrow~ 0 \ ,\quad \tan (\theta) ~\longrightarrow~ \tan(\theta_{\rm W})\ ,
\eeqn
i.e.  $\theta$ becomes the weak angle, and the other angles
vanish. In this limit the mass $M_-$ takes the standard expression
for the mass of the Z-boson,
\beqn
M_{{\rm Z}'}^2 ~\longrightarrow~ M^2 \ , \quad M_{\rm Z}^2 ~\longrightarrow~ \frac14 v^2( g_2^2 + g_Y^2 ) \ ,
\eeqn
the mass squared matrix being block-diagonal. Remarkably, in the
limit $v/M\rightarrow  0$, with $\delta$ fixed, which corresponds
to a large overall mass scale compared to the electro-weak scale
set by the Higgs expectation value, only one of the angles
vanishes,
\beqn
\tan (\psi) ~\longrightarrow~ 0 \ .
\eeqn
The other two parameters are independent of $v^2$. No matter how
high the scale would be, at which the additional couplings are
generated, the low energy parameters $\tan (\phi)$ and
$\tan(\theta)$ can deviate from SM expressions, since deviations
are not suppressed by the high scale.
\\

For the purpose of obtaining a more physical parametrization, it
is useful to replace the parameters $v, vg_2, vg_Y, M_1, M_2$ of
the Stueckelberg extended model with those of the SM Lagrangian,
and fix them through measured quantities, up to the mass scale
$M$. Defining
\beqn
M^2_Y = \frac14 g_Y^2 v^2 = \frac{\frac14 (ev)^2 M_{\rm W}^2}{M_{\rm W}^2-\frac14 (e v)^2} \ ,
\eeqn
where $v = (\sqrt{2}G_F)^{-1/2}$,
we can express $M_1$ and $M_2$, or $\delta$ and $M$,
in terms of $M_{\rm Z}, M_{{\rm Z}'},M_{\rm W} , M_Y$ via
\beqn \label{m2m1}
M_1^2 &=&
\frac{M_{{\rm Z}'}^2 (M_{\rm Z}^2 - M_{\rm W}^2) + M_{\rm W}^2 (M_{\rm W}^2 +M_{Y}^2 -M_{\rm Z}^2)}
     {M_{Y}^2} \ ,
\non
M_2^2 &=&
\frac{(M_{{\rm Z}'}^2 -M_{\rm W}^2 -M_{Y}^2)(M_{\rm Z}^2 - M_{\rm W}^2 -M_{Y}^2))}{M_Y^2}  \ ,
\non
\frac{M^2_2}{M^2_1} &=& \delta^2 ~=~
\frac{( M_{\rm Z}^2 - M_{\rm W}^2 - M_Y^2 )( M_{{\rm Z}'}^2 - M_{\rm W}^2 - M_Y^2 )}
     {M_{{\rm Z}'}^2 (M_{\rm Z}^2 - M_{\rm W}^2) + M_{\rm W}^2 (M_{\rm W}^2 +M_{Y}^2 -M_{\rm Z}^2)} \ .
\eeqn
Now the Stueckelberg Lagrangian is fixed by adjusting the
parameters to fit  the experimental parameters. This requires
global fits to the electro-weak data which is outside the scope of
this work. If implemented, it should determine the full allowed
range of the \st parameter space in $M_1, M_2$. To illustrate the
typical values, one convenient choice is to pick $M_{{\rm Z}'}$
and $\delta$. Once these are fixed, one can compute the three
angles, $\theta, \phi, \psi$. For instance, for
\beqn
\delta ~=~ 0.029\ , \quad
M_{{\rm Z}'} ~=~ 250\, {\rm GeV}\
\eeqn
we find
\beqn
\tan(\phi) = 0.029 \ , \quad
\tan(\psi) = 0.002\ , \quad
\tan(\theta) = 0.546 \ .
\eeqn
Note that characteristically, $|\psi|\sim \frac{1}{10}|\phi|$ and
$\theta$ equals $\theta_W$ up to less than a percent.


\subsection{Couplings to fermions}

Defining a vector of neutral currents $J^\mu_a = (g_X J_X^\mu, g_Y
J_Y^\mu, g_2 J^{3 \mu}_2 )$, the couplings to the fermions are
easily found by inserting the mass eigenstates into the neutral
current (NC) interaction Lagrangian
\beqn
{\cal L}_{\rm NC} = g_2 A_\mu^3 J^{3\mu}_2 + g_Y B_\mu J^\mu_Y + g_X C_\mu J^\mu_X =
\sum_{a=1}^3 V_{\mu a} J^{\mu}_a
= \sum_{a,b=1}^3 E^{[1]}_{\mu a} {\cal O}^{[1]}_{ba} J^{\mu}_b
\ .
\eeqn
The three components of this interaction product are easily
expressed through the angle parameters,
\beqn
\sum_{b=1}^3 {\cal O}^{[1]}_{ba} J^{\mu}_b = \left[
\begin{array}{c}
\frac{\sin(\psi)}{\sqrt{g_2^2 + g_Y^2\cos^2(\phi)}} \left(
   \cos^2(\phi) g_Y^2 J_Y^\mu - g_2^2 J_2^{3\mu} - \frac12 \sin(2\phi) g_Xg_Y J_X^\mu \right)
\\
   + \cos(\psi) ( \sin(\phi) g_Y J_Y^\mu + \cos(\phi) g_X J_X^\mu)
\\[.5cm]
\frac{\cos(\psi)}{\sqrt{g_2^2 + g_Y^2\cos^2(\phi)}} \left(
   \cos^2(\phi) g_Y^2 J_Y^\mu - g_2^2 J_2^{3\mu} - \frac12 \sin(2\phi) g_Xg_Y J_X^\mu \right)
\\
- \sin(\psi) ( \sin(\phi) g_Y J_Y^\mu + \cos(\phi) g_X J_X^\mu )
\\[.5cm]
\frac{g_2g_Y\cos(\phi)}{\sqrt{g_2^2+g_Y^2 \cos^2(\phi)}}
  \left( J_Y^\mu + J_2^{3\mu} - \frac{g_X}{g_Y} \tan(\phi) J^\mu_X \right)
\end{array} \right]\ .
\eeqn
The first line couples to Z$'$, the second to Z, and the third line is the modified
electromagnetic current.
\\

The modification of the current that couples to the photon leads
to two effects: First the electric charges of the fields of the SM
would get modified. For instance, the charge of the up and the
down quark are
\beqn
Q_u= \frac{2}{3} - \frac{g_X}{g_Y} \tan(\phi) Q_X(u) \ ,\quad
   Q_d= -\frac{1}{3} - \frac{g_X}{g_Y} \tan(\phi) Q_X(d)\ .
\label{qu}
\eeqn
However, the charge neutrality of the neutron requires that
$Q_u+2Q_d=0$ to very high precision. This, and similar relations
for all other fields of the SM, can only be satisfied if the
$U(1)_X$ charges were proportional to their electric charges
or vanishing. We, therefore, make
the assumption that all fields of the SM itself are neutral under
the extra $U(1)_X$ gauge symmetry, setting $Q_X({\rm SM})=0$. This
means that the couplings of $C_{\mu}$ with visible matter are
strictly forbidden, in order to maintain the charge cancellation
between quarks or leptons. On the other hand, there is a priori no
such restriction on the matter in the hidden sector to which
$C_{\mu}$ can couple. This implies that the masses of charged
matter fields in the hidden sector have to be safely outside the
current limits of direct detection.
\\

Second, there still is a modification of the electric charge
$e$, the coupling that appears in the term
\beqn
e A_\mu^\gamma J^\mu_{\rm em} = e A_\mu^\gamma \big( J_Y^\mu + J_2^{3\mu} \big) \ ,
\eeqn
which is now defined by
\beqn
e = \frac{g_2g_Y\cos(\phi)}{\sqrt{g_2^2+g_Y^2 \cos^2(\phi)}} \ .
\eeqn
Thus, the Stueckelberg mechanism effectively changes $g_Y$ to
$g_Y\cos(\phi)$. All these modifications, of course, go away, when
one takes the SM limit $\delta \rightarrow 0$, when
$\cos(\phi)\rightarrow 1$. Similarly, the standard coupling of the
Z-boson,
\beqn
{\rm Z}_\mu g_{\rm NC} J^\mu_{\rm NC} ~\longrightarrow~ {\rm Z}_\mu \frac{1}{\sqrt{g_2^2 + g_Y^2}} \left(
   g_Y^2 J_Y^\mu - g_2^2 J_2^\mu \right)
\eeqn
is recovered in this limit. One can also read off that the angle
$\psi$ takes the role of mixing the couplings of Z and Z$'$. An
important feature of this interaction Lagrangian is that the
coupling constants of the extra gauge boson are not arbitrary
parameters, but uniquely defined through $\delta$ and $M$, the
only new parameters of the model (aside from $g_X$, which we
always assume to be of the same order as $g_Y$ or $g_2$). We
postpone a discussion of more experimental properties and concrete
signatures of the StSM for later, when we treat the supersymmetric
and the ordinary Stueckelberg extension in a combined fashion.


\section{The \st extension of MSSM}

In this section we give the \st extension of the minimal
supersymmetric standard model (MSSM) \cite{kn2} which may be
labelled the StMSSM. The gauge symmetry is again extended by a
single abelian factor $U(1)_X$, and only the neutral interactions
are affected by the Stueckelberg mechanism. As argued above, we
now assume that the fields of the MSSM are neutral under the new
$U(1)_X$. Since supersymmetry requires the extra fields to fall
into proper multiplets, we  add one chiral (or linear) and one
vectorsupermultiplet to the MSSM, which combine into a massive
spin one multiplet and mix with the other massive vector
multiplets after the condensation of the Higgs boson. Beyond the
\st chiral and vector superfields we in principle also allow for
the existence of a hidden sector.
\\

In setting up the supersymmetric extension (using standard
superspace notation \cite{WessBagger}) we consider the following
action for the Stueckelberg chiral multiplet
$S=(\rho+i\sigma,\chi,F_S)$
\beqn
{\cal L}_{\rm St} = \int d^2\te d^2\thb\ (M_1C+M_2B+  S +\bar S )^2\ ,
\label{mass}
\eeqn
where $C=(C_\mu,\lambda_C,D_C)$ is the gauge vectormultiplet for
$U(1)_X$, $B$ that for the hyper charge. The supersymmetrized
gauge transformations under the new $U(1)_X$ are
\beqn \label{stgauge}
\delta_Y B = \Lambda_Y + \bar\Lambda_Y  \ , \quad
\delta_Y S = - M_2 \Lambda_Y\ ,
\eeqn
and for the hyper charge
\beqn
\delta_X C = \Lambda_X + \bar\Lambda_X  \ , \quad
\delta_X S = - M_1 \Lambda_X\ .
\eeqn
Although $S$ transforms under the abelian gauge symmetries, it is
somewhat misleading to think of it as a charged field in the
standard sense of a charged chiral multiplet. To be slightly more
specific on our notation, we denote $C$ by
\beqn
C~=~ -\theta\sigma^{\mu}\bar \theta C_{\mu}
+i\theta\theta \bar\theta \bar \lambda_C
-i\bar\theta\bar\theta \theta  \lambda_C
+\frac{1}{2} \theta \theta\bar\theta\bar\theta D_C\ .
\eeqn
Similarly for $B$ with $B_\mu, \lambda_B$ and $D_B$, and
$S$ is given by
\beqn \label{superS}
S &=& \frac{1}{2}(\rho +i\sigma ) + \theta \chi
 + i \theta\sigma^{\mu}\bar\theta \frac{1}{2}(\partial_{\mu} \rho
+i \partial_{\mu} \sigma) \nonumber\\
&&
+ \theta\theta F_S + \frac{i}{2} \theta \theta \bar\theta \bar\sigma^{\mu} \partial_{\mu}\chi
+\frac{1}{8}\theta\theta\bar\theta\bar\theta (\Box \rho+i\Box
\sigma)\ .
\eeqn
Its scalar component contains the scalar $\rho$ and the axionic
pseudo-scalar $\sigma$. This leads to \cite{fayet,kleinetal}
\beqn \label{stueck}
{\cal L}_{\rm St} &=& - \frac{1}{2}(M_1C_{\mu} +M_2 B_{\mu} +\partial_{\mu} \sigma)^2
 - \frac{1}{2} (\partial_\mu \rho)^2
- i \chi \sigma^{\mu} \partial_{\mu}\bar \chi +2|F_S|^2
\\
&&  
 +\rho(M_1D_C +M_2 D_B)
 +\big [ \chi (M_1 \lambda_C + M_2 \lambda_B)
 + {\rm h.c.} \big]\ .
\nonumber
\label{ls}
\eeqn
For the gauge fields we add the standard kinetic terms
\beqn \hspace{-.5cm}
{\cal L}_{\rm gkin} &=&
-\frac{1}{4} C_{\mu\nu} C^{\mu\nu} -\frac{1}{4} B_{\mu\nu} B^{\mu\nu} -
i \lambda_B\sigma^{\mu}\partial_{\mu} \bar \lambda_B
-i \lambda_C\sigma^{\mu}\partial_{\mu} \bar \lambda_C
+\frac{1}{2} D_C^2  +\frac{1}{2} D_B^2\ .
\nonumber
\eeqn
For the matter fields, chiral superfields $\Phi_i$ and $\Phi_{{\rm
hid},i}$ are introduced. The fermions (quarks $q_i$, leptons
$l_i$, Higgsinos $\tilde h_i$) of the MSSM will be collectively
denoted as $f_i$, hidden sector fermions as $f_{{\rm hid}, i}$.
The scalars (sfermions $\tilde q_i$, sleptons $\tilde l_i$ and the
two Higgs fields $h_i$) are summarized as $z_i$ and $z_{{\rm
hid},i}$.\footnote{The matter chiral multiplets are defined
exactly according to the conventions of \cite{WessBagger}, while
$S$ carries some extra factors for convenience.} The Lagrangian
reads
\beqn
{\cal L}_{\rm matt} ~=~ \int d^2\te d^2\thb\, \Big[
\sum_i \bar \Phi_i e^{2g_Y Q_Y B+ 2g_X Q_X C} \Phi_i
 + \sum_i \bar \Phi_{{\rm hid},i} e^{2g_Y Q_Y B+ 2g_X Q_X C} \Phi_{{\rm hid},i}\Big] \ .
\nonumber
\eeqn
where $Q_Y=Y/2$, and where $Y$ is the hyper charge so that
$Q=T_3+Y/2$. As mentioned already, the SM matter fields do not
carry any charge under the hidden gauge group, i.e. $Q_X \Phi_i
=0$. Thus we have
\beqn \label{matt}
{\cal L}_{{\rm matt},i} &=&
- |D_\mu z_i|^2 - i f_i \sigma^\mu \partial_\mu \bar f_i + |F_i|^2
+ g_Y B_\mu J_{Yi}^\mu + g_X C_\mu J_{Xi}^\mu
\\
&& \hspace{-1.5cm}
- \sqrt{2} \big[ i g_Y Q_Y z_i \bar f_i \bar \lambda_B + i g_X Q_X z_i \bar f_i \bar \lambda_C
                   + {\rm h.c.} \big]
+ g_Y D_B (\bar z_i Q_Y z_i) + g_X D_C (\bar z_i Q_X z_i) \ ,
\nonumber
\eeqn
where $D_\mu  = \partial_\mu + ig_Y Q_Y B_\mu + ig_X Q_X C_\mu$, and
\beqn
J_{Yi}^\mu = f_i Q_Y \sigma^\mu \bar f_i \ , \quad
J_{Xi}^\mu = f_i Q_X \sigma^\mu \bar f_i \ .
\eeqn
The above uses standard notation with Weyl spinors. It is
convenient before passing to mass eigenstates to define now
Majorana spinors in the form
\beqn
\psi_S~=~
\left(\matrix{\chi_{\alpha}\cr
\bar \chi^{\dot{\alpha}}}\right)\ ,~~~~
\lambda_X~=~
\left(\matrix{\lambda_{C\alpha}\cr
\bar \lambda^{\dot{\alpha}}_C}\right)\ ,~~~~
\lambda_Y~=~
\left(\matrix{\lambda_{B\alpha}\cr
\bar \lambda^{\dot{\alpha}}_B}\right)\ .
\eeqn
Thus the \st extension introduces two new Majorana spinors in the
system, i.e. $\psi_S$ and $\lambda_X$. We also rewrite the matter
fermions in terms of Majorana fields, but still use the same
symbols $f_i$ here, as before for the Weyl fermions. One has for
instance the following identities
\beqn
\chi\lambda_C +\bar\chi\bar\lambda_C &=& \bar\psi_S\lambda_X\ ,
\non
\chi\lambda_C - \bar \chi \bar\lambda_C &=& \bar \psi_S \gamma_5 \lambda_X \ ,
\non
\chi \sigma^{\mu}\partial_{\mu}\bar \chi -
(\partial_{\mu}\chi) \sigma^{\mu}\bar \chi &=&
\bar\psi_S \gamma^{\mu}\partial_{\mu}\psi_S\ .
\eeqn
We may then write the total Lagrangian (by substituting back in
the values for $D_B$ and $D_C$) in the form
\beqn \label{stmssm}
{\cal L}_{\rm St}+{\cal L}_{\rm gkin}+{\cal L}_{{\rm matt},i} &=& \non
&& \hspace{-4.5cm}
-\frac{1}{2}(M_1C_{\mu} +M_2 B_{\mu} +\partial_{\mu} \sigma)^2
-\frac{1}{2} (\partial_\mu \rho)^2 - \frac12 ( M_1^2 + M_2^2) \rho^2
- \frac{i}{2} \bar\psi_S \gamma^{\mu} \partial_{\mu} \psi_S
\non
&&  \hspace{-4.5cm}
-\frac14 B_{\mu\nu}B^{\mu\nu}
-\frac14 C_{\mu\nu}C^{\mu\nu}
-\frac{i}{2} \bar\lambda_Y \gamma^\mu \partial_\mu \lambda_Y
-\frac{i}{2} \bar\lambda_X \gamma^\mu \partial_\mu \lambda_X
- |D_\mu z_i|^2 - \frac{i}{2} \bar f_i \gamma^\mu \partial_\mu f_i
\nonumber\\
&&  \hspace{-4.5cm}
+ \frac12 g_Y B_\mu  \bar f_i \gamma^\mu Q_Y f_i
+ \frac12 g_X C_\mu  \bar f_i \gamma^\mu Q_X f_i
+ M_1 \bar \psi_S\lambda_X + M_2\bar\psi_S\lambda_Y
\nonumber\\
&&  \hspace{-4.5cm}
- \sqrt 2 g_X \big[ i z_i Q_X \bar f_i \lambda_X + {\rm h.c.} \big]
- \rho \Big( g_Y M_2 ( \bar z_i Q_Y z_i ) + g_X M_1 ( \bar z_i Q_X z_i ) \Big)
\nonumber\\
&&  \hspace{-4.5cm}
-\frac12  \Big[ \sum_i  \bar z_i g_Y Q_Y z_i
\Big]^2 -\frac12 \Big[ \sum_i \bar z_i g_X Q_X z_i \Big]^2 \ .
\eeqn
Of course, one has to add the hidden sector fields and sum over
$i$ when appropriate. We have already pointed out that the  MSSM
itself is neutral under the new gauge symmetry $U(1)_X$, but the
hidden sector fields may well be charged under it. In order to get
a model that represents the pure \st effect, we further let the
hidden sector be neutral under the gauge group of the SM, i.e.\ we
really demand it to be hidden with respect to the SM gauge
interactions.
\\

The modifications that are introduced by the \st extension are now
completely evident: We have added the degrees of freedom of one
abelian gauge vector multiplet, the vector field $C_\mu$ and its
gaugino $\lambda_X$, as well as the chiral multiplet with the
complex scalar $\rho+ia$ and the fermion $\psi_S$. There are three
channels for the new sector to communicate to the SM fields: $i)$
the mixing of neutral gauge bosons through the non-diagonal vector
boson mass matrix, just as in the Stueckelberg extension of the
SM, $ii)$ the mixing of neutralinos through the fermion mass
matrix with the off-diagonal terms involving the gauginos and
$\psi_S$, $iii)$ the cubic couplings of $\rho$ with the scalar
partners of SM fermions and the Higgs bosons.
\\

Through the Stueckelberg coupling, a combination of the vector
fields $B_\mu$ and $C_\mu$ gets a mass, and absorbs the axionic
component $\sigma$ as its longitudinal mode. The real part $\rho$
gets a mass $M$. We shall see that mass eigenstates that combine
out of the two gauginos and $\psi_S$ will just form a massive
fermion of identical mass as the vector and the scalar. Thus, out
of the massless two vector and one chiral multiplet, one massive
spin one (out of a vector, a Dirac fermion and a scalar) and one
massless vector multiplet are combined. When the Higgs condensate
is introduced, the massless vector multiplet will mix with the
3-component of the adjoint $SU(2)_L$ gauge boson multiplet.


\subsection{Adding soft supersymmetry breaking terms}

Including soft supersymmetry breaking terms will finally break up the
mass degeneracy of the spectrum.
The soft breaking terms  relevant for the further discussion are
\beqn
{\cal L}_{\rm soft} &=&
-\frac12 m_\rho^2 \rho^2
-\frac12 \tilde m_Y \bar \lambda_Y \lambda_Y
-\frac12 \tilde m_X \bar \lambda_X \lambda_X
\non
&&
- m_1^2 |h_1|^2
- m_2^2 |h_2|^2 - m_3^2 ( h_1 \cdot h_2 + \ {\rm h.c.}\ ) \ ,
\eeqn
with $m_1^2= m_{h_1}^2+|\mu|^2$, $m_2^2= m_{h_2}^2+|\mu|^2$,
$m_3^2=|\mu B|$, where $\mu$ is the Higgs mixing parameter (which
is not really soft but part of the superpotential) and $B$ is the
soft bilinear coupling. Note that there is no soft mass for the
chiral fermion $\psi_S$.


\subsection{Adding Fayet-Illiopoulos terms}

The above analysis was so far without the Fayet-Illiopoulos
(FI) terms. In the present case it means that one has the freedom
to introduce two terms in the Lagrangian  of the form
\beqn
{\cal L}_{\rm FI} &=& \xi_B D_B + \xi_C D_C\ .
\eeqn
For the contribution of $\xi_B$ we make the usual assumption that
it is subdominant and can be neglected in the Higgs potential that
drives spontaneous gauge symmetry breaking. This remains true for
the modified field
\beqn
- D_B =  \xi_B + M_2 \rho + g_Y \sum_i \bar z_i Q_Y z_i \ ,
\eeqn
as it will turn out that the modification $M_2 \rho$ will be very small.
For the FI-term with $\xi_C$ one finds on eliminating
the auxiliary field $D_C$
\beqn
- D_C =  \xi_C + M_1 \rho + g_X \sum_i \bar z_i Q_X z_i \ .
\eeqn
The modification of Eq.(\ref{stmssm}) in  the presence of FI terms
is implemented by the  replacement
\beqn
\sum_i  \bar z_i g_Y Q_Y z_i &\rightarrow&
\xi_B + \sum_i  \bar z_i g_Y Q_Y z_i \ ,
\nonumber\\
\sum_i \bar z_i g_X Q_X z_i &\rightarrow&
\xi_C + \sum_i \bar z_i g_X Q_X z_i 
\eeqn
in Eq.(\ref{stmssm}). Since we assume that the charges $Q_X$ of
the MSSM fields are all vanishing, this will not have any impact
on the visible sector mass or quartic couplings. Depending on the
charges of the hidden sector field, such a FI-term may be able to
drive a spontaneous breaking of the $U(1)_X$ gauge symmetry, which
would result in a mass term for the photon mass eigenstate, and
thus has to be excluded.


\section{Implications and Predictions}

Here we now discuss the consequences of the extensions of the SM
or MSSM with an extra $U(1)_X$ that couples to a pseudo-scalar
$\sigma$, together with the hyper charge gauge boson multiplet, in
the way of the Stueckelberg mechanism. First, we shall go through the modifications
of the SM. They all refer to the non-diagonal mass squared matrix
of the neutral gauge bosons, and the effects of their mixing.
These effects will also be reproduced in the MSSM without any
modification, which then contains further signatures through the
modified neutral scalar and neutral fermion sectors.


\subsection{Comparison to the Standard Model}

Some of the implications of the extended model have already been
explained above. Roughly speaking, diagonalizing the mass squared
matrix of the neutral gauge bosons introduces a mixing of all
three vector fields, and of the currents they couple to. For the
photon this implies that the coupling constant to the
electromagnetic current is modified, and that it may couple to
hidden sector matter charged under $U(1)_X$. The latter is a very
interesting phenomenon, since it may give indirect evidence of
hidden sector matter, which is otherwise invisible to gauge
interactions. However, the couplings to the hidden sector are
highly model dependent and could even be completely suppressed as
discussed at the end of section 7. For the neutral current
interactions, the mixing also implies a change of coupling
constants and currents, and a coupling to hidden matter. The
latter may not be so dramatic here, since the interactions are
only short-ranged.


\subsubsection{Neutral current interactions: $\rho$ parameters}

A useful parameter to study the neutral current interactions is
the conventional $\rho$ parameter which is defined as the ratio in
the effective low energy Lagrangian of the neutral and the charged
current interactions.\footnote{The $\rho$ parameters discussed
here should not be confused with the scalar field $\rho$ that
appears in the \st extension of the MSSM.} For the SM at the tree
level this ratio is
\beqn
\rho_{\rm SM}= \frac{M_{\rm W}}{\cos(\theta_W) M_{\rm Z}}\ ,
\eeqn
and there are small deviation from unity due to radiative
corrections. For the model at hand this issue is more complicated,
and the neutral and charged current interactions can no longer be
compared with just one ratio, because there are now two neutral
massive gauge bosons. To see this, we can eliminate Z- and
Z$'$-bosons at low energy to obtain an effective neutral current
interaction, which we can write as follows
\beqn
{\cal L}_{\rm NC-eff} &=&
\frac{4G_F}{\sqrt 2} 2 \Big[
 \rho_{\rm Z} (J^{\mu}_3-\sin^2(\theta_{\rm Z}) J^{\mu}_{\rm em})
 (J_{3\mu}-\sin^2(\theta_{\rm Z}) J_{{\rm em} \mu})
\nonumber\\
&& \hspace{1.5cm}
+ \rho_{{\rm Z}'} (J^{\mu}_3-\sin^2(\theta_{{\rm Z}'}) J^{\mu}_{\rm em})
(J_{3\mu}-\sin^2(\theta_{{\rm Z}'}) J_{{\rm em} \mu}) \Big] \ ,
\label{ncurrent}
\eeqn
while the charged effective current-current Lagrangian is unchanged
\beqn
{\cal L}_{\rm CC-eff} &=&
\frac{4G_F}{\sqrt 2} J_\mu^+ J^{- \mu} \ .
\eeqn
Above, we have defined $\rho_{\rm Z}$ and $\rho_{{\rm Z}'}$ by
\beqn
\rho_{\rm Z} =
\frac{M_{\rm W}^2 f_{\rm Z}^2}{M_{\rm Z}^2 \cos(\theta)}
\ , \quad
\rho_{{\rm Z}'}= \frac{M_{\rm W}^2 f_{{\rm Z}'}^2}{M_{{\rm Z}'}^2 \cos(\theta)}
\eeqn
and the effective decay constants $f_{\rm Z}$ and $f_{{\rm Z}'}$ by
\beqn
f_{\rm Z} = \cos(\psi) +\sin(\theta)\tan(\phi) \sin(\psi)
\ , \quad
f_{{\rm Z}'} = \sin(\psi) +\sin(\theta)\tan(\phi) \cos(\psi)\ .
\eeqn
Finally $\theta_{\rm Z}$ and $\theta_{{\rm Z}'}$ are defined by
\beqn
\sin^2(\theta_{\rm Z}) &=&
\frac{\sin^2(\theta) -\sin(\theta) \tan(\phi) \tan(\psi)}
{1+\sin(\theta) \tan(\phi) \tan(\psi)} \ ,
\nonumber\\
\sin^2(\theta_{{\rm Z}'}) &=&
\frac{\sin^2(\theta)\tan(\psi) + \sin(\theta) \tan(\phi)}
{\tan(\psi)+\sin(\theta) \tan(\phi)} \ .
\eeqn
Further we can also define a parameter $\rho$ analogous to the
conventional parameter in the SM so that $\rho = M_{\rm W}/(M_{\rm
Z}\cos(\theta))$. Even at the tree-level $\rho_{\rm Z}, \rho_{{\rm
Z}'}$, and $\rho$ are all different. Further, Eq.(\ref{ncurrent})
shows that in the present model different combinations of
$\rho_{\rm Z}$ and $\rho_{{\rm Z}'}$ appear for the operators
$J^{3}_2 \cdot J^3_{2}$, $J_{\rm em} \cdot J_{{\rm em}}$, and
$J^{3}_2 \cdot J_{{\rm em}}$. Thus the relevant ratio of charged
and neutral interaction strength will depend on the process. In
the limit that $M_2=0$ one has $\rho_{{\rm Z}'}=0$ and $\rho_{\rm
Z}=\rho=\rho_{\rm SM}$.
\\

Currently, there are stringent constraints on the neutral current
processes and the data is consistent with the SM. However, the
error corridor in the experimental measurements allow the
possibility of new physics including the possibility of new Z$'$
bosons and this topic has been investigated extensively in the
literature. This possibility also applies to the current model if
the contribution of the new sector is sufficiently small to be
consistent with the experimental error corridor. Thus, for
example, for $\psi \sim 1^0, \phi\sim 1^0$, and setting
$\sin^2(\theta)=\sin^2(\theta_W) =0.23$ one finds,
$\sin^2(\theta_{\rm Z})=0.2298$, and $1-\rho_{\rm Z}/\rho_{\rm
SM}=0.0001$, while $\rho_{{\rm Z}'}/\rho_{\rm SM}=0.025 \times M_{\rm
Z}^2/M_{{\rm Z}'}^2$, which gives $\rho_{{\rm Z}'}/\rho_{\rm
SM}\sim 0.0025$ for $M_{{\rm Z}'}/M_{{\rm Z}}=3$. These are
consistent with the current error corridors on $\rho$, of
the order of $0.005$.


\subsubsection{Visible width and branching ratios of Z$'$}

In greater detail, one can write the couplings of the first
generation as follows
\beqn
{\cal L}_{{\rm Z}l \bar l} &=&
-\frac{1}{2} \Big[ - \frac{g_2^2-\cos^2(\phi) g_Y^{2}}{\sqrt{g_2^2+\cos^2(\phi) g_Y^{2}}}
\cos\psi -
\sin(\phi)\sin(\psi) g_Y  \Big] \, \bar e_L\gamma^{\mu}e_L {\rm Z}_{\mu}
\nonumber\\
&&
- \Big[ \frac{\cos^2(\phi) g_Y^2 }{\sqrt{g_2^2+\cos^2(\phi) g_Y^2 }} \cos(\psi)
 - \sin(\phi) \sin(\psi) g_Y \Big]\, \bar e_R\gamma^{\mu} e_R {\rm Z}_{\mu}
\nonumber\\
&&
- \frac{1}{2} \Big[
\sqrt{g_2^2+\cos^2(\phi) g_Y^2 } \cos(\psi)
- \sin(\phi) \sin(\psi) g_Y  \Big] \, \bar \nu_e \gamma^{\mu}\nu_e
{\rm Z}_{\mu}
\nonumber\\
&&
-\frac{1}{2} \Big[ -\frac{g_2^2-\cos^2(\phi) g_Y^2 }{\sqrt{g_2^2+\cos^2(\phi) g_Y^2 }}
\sin(\psi) + \sin(\phi) \cos(\psi) g_Y\Big] \, \bar e_L\gamma^{\mu}e_L  {\rm Z}_{\mu}'
\nonumber\\
&&
-\Big[  \frac{\cos^2(\phi) g_Y^2 }{\sqrt{g_2^2+\cos^2(\phi) g_Y^2 }} \sin(\psi) +
\sin(\phi) \cos(\psi) g_Y \Big] \, \bar e_R\gamma^{\mu} e_R {\rm Z}_{\mu}'
\nonumber\\
&&
-\frac{1}{2} \Big[ \sqrt{g_2^2+\cos^2(\phi) g_Y^2 }\sin(\psi) +
\sin(\phi) \cos(\psi) g_Y  \Big] \, \bar \nu_e \gamma^{\mu}\nu_e{\rm Z}_{\mu}' \ .
\label{zzp1}
\eeqn
And the couplings of Z and Z$'$ with quarks are given by
\beqn
{\cal L}_{{\rm Z}q\bar q} &=& - \sqrt{g_2^2+\cos^2(\phi) g_Y^2}
\\
&& \hspace{-.5cm} \times \Big[
{\rm Z}_{\mu}\left( J_3^{\mu} -\sin^2(\theta)
J_{\rm em}^{\mu} \right) \cos(\psi)
+{\rm Z}_{\mu}(J_{\rm em}^{\mu}-J_3^{\mu}) \sin(\theta)
\tan(\phi) \sin(\psi)
\nonumber\\
&&
+{\rm Z}_{\mu}'(J_3^{\mu} -\sin^2(\theta) J_{\rm em}^{\mu})\sin(\psi)
-{\rm Z}_{\mu}' (J_{\rm em}^{\mu}-J_3^{\mu}) \sin(\theta)
\tan(\phi) \cos(\psi)
\Big]
\nonumber \ .
\label{zzp2}
\eeqn
In addition to the new couplings of the quarks to the ${\rm
Z}_{\mu}'$ boson the couplings of ${\rm Z}_{\mu}$ with quarks are
also affected. Below we give a comparison of the decay branching
ratios for the decay of the ${\rm Z}'$ into quarks and leptons
versus the branching ratios for the decay of the Z into quarks and
leptons. We display the results for $|\psi|\ll |\phi|$ in Table 1.
\begin{table}[h]
\begin{center}
%
\begin{tabular}{|c|c|c|}
\hline
&& \\[-.35cm]
Ratio of branching ratios & Z decay & Z$'$ decay \\[.1cm]
\hline\hline
&& \\[-.4cm] ${\it l}\bar{\it l}/\nu\bar\nu$      &  0.5             &   5       \\[.05cm]
\hline
&& \\[-.4cm] $b\bar b/\tau\bar \tau$  &  $(3-4s_W^2+\frac{8}{3}s_W^4)/(1-4s_W^2+8s_W^4)$   &   $\frac{1}{3}$ \\[.05cm]
\hline
&& \\[-.4cm]  $u\bar u/d \bar d$
  & $(3-8s_W^2+\frac{32}{3}s_W^4)/(3-4s_W^2+\frac{8}{3}s_W^4)$ & $\frac{17}{5}$      \\[.05cm]
\hline
\end{tabular}
\caption{A comparison of the ratio of branching ratios into quarks
and leptons ${\rm Z}'$ versus Z ($s_W=\sin(\theta_W)$).}
\end{center}
\end{table}

In the same approximation the total decay width of ${\rm Z}'$ into
the visible sector quarks and leptons
is given by
\beqn
\Gamma({\rm Z}'\rightarrow \sum_i f_i\bar f_i)\simeq
M_{{\rm Z}'} g_Y^2\sin^2(\phi) \times \left\{ {\frac{103}{288\pi}~~ {\rm for}\ M_{{\rm Z}'} < 2m_t
                  \atop  \frac{5}{12\pi}~~ {\rm for}\ M_{{\rm Z}'} > 2m_t  }\right.
\eeqn
The decay signatures of the ${\rm Z}'$ boson are very different
from those of the Z boson of the SM. The reason for this
difference arises from the fact that the ${\rm Z}'$ dominantly
decays via the couplings proportional to $g_Y$ as can be seen by
making  the approximation $|\psi|\ll |\phi|\ll 1$ in
Eq.(\ref{zzp1}) and Eq.(\ref{zzp2}).


\subsection{The bosonic sector of the extended MSSM}

The bosonic sector of the StMSSM consists of the neutral vector
bosons, the Stueckelberg scalar $\rho$, the Higgs fields and the
sfermions of the MSSM. The Stueckelberg axion $\sigma$ is decoupled
after gauge fixing, and is absorbed by the gauge bosons. The
analysis of the mass matrix of the vector bosons remains unchanged
from that of the SM as discussed in section 2 and we
do not have to repeat it here.
\\

We have already mentioned the assumptions that go into the
definitions of the model. We take all the matter fields of the
MSSM and the two Higgs multiplets to be neutral under the
$U(1)_X$, and we also demand that there is no charged scalar
condensate formed in the hidden sector, e.g.\ no vacuum
expectation value $\langle \bar z_i Q_X z_i\rangle \not= 0$. This
would add another term to the mass matrix (\ref{vecmass}) and
finally give a mass to the photon eigenstate. We, therefore, impose
$\langle z_i\rangle=0$ for all hidden scalars $z_i$ that carry
charge under $U(1)_X$, which are the only ones relevant for us.
\\

Under these assumptions, the subsector of the StMSSM which
contains the neutral vector bosons, and their couplings to the
conserved currents is just identical to the StSM. We are left in
the bosonic sector with the extra neutral scalar $\rho$, that
mixes with the neutral components of the Higgs doublets.
\\


\subsubsection{The scalar Higgs fields and the \st scalar $\rho$}

The scalar potential for the two Higgs-doublets of the MSSM plus
the \st scalar $\rho$ involves a non-diagonal mass squared matrix,
similar to the mixing of neutral gauge bosons. As explained in the
previous section, the Higgs fields are neutral under $U(1)_X$,
hidden sector fields are neutral under hyper charge, and there are
no condensates charged under $U(1)_X$. Then we get
\beqn
{\cal V}(h_1,h_2,\rho) &=&
 ( m_1^2 - \frac{1}{2}\rho g_Y M_2 ) |h_1|^2
+   (m_2^2 + \frac{1}{2}\rho g_Y M_2 ) |h_2|^2
+ m_3^2 (h_1 \cdot h_2 +\ {\rm h.c.}\ )
\nonumber\\
&&
+\frac{g_2^2+g_Y^2}{8} |h_1|^4 + \frac{g_2^2+g_Y^2}{8} |h_2|^4
+ \frac{g_2^2-g_Y^2}{4} |h_1|^2 |h_2|^2 -\frac{g^2_2}{2} |h_1 \cdot h_2|^2
\nonumber\\
&&
+\frac{1}{2} (M_1^2+M_2^2+m_{\rho}^2) \rho^2\ .
\eeqn
The Higgs doublets are defined $h_1 = (h_1^0 , h_1^-)^T$, $h_2=(h_2^+,h_2^0)^T$, and
$h_1\cdot h_2 = h_1^0 h_2^0 - h_1^- h_2^+$.
For the Higgs scalars $h_1^0$, $h_2^0$, and  for
$\rho$
we make replacements
\beqn
h_1^0 \rightarrow \frac{1}{\sqrt 2}(v_1+h^0_1)\ , \quad
h_2^0 \rightarrow \frac{1}{\sqrt 2}(v_2+h^0_2)\ , \quad
\rho \rightarrow  v_\rho + \rho \  ,
\eeqn
where $v_i$ and $v_\rho$ are the vacuum expectation values, and
assumed to be real. As usual, they are parameterized by
\beqn
v_1 = v \cos (\beta) \ , \quad
v_2 = v \sin(\beta) \ .
\eeqn
For $\rho$ one has
\beqn
v_\rho = \frac{2g_Y M_{\rm W}^2 M_2}{g_2^2 M_\rho^2}  \cos(2\beta) \ ,
\eeqn
where $M_\rho^2 = M^2 + m_\rho^2= M_1^2 +M_2^2 + m_\rho^2$.
To give a rough estimate for large $\tan(\beta)$, one has
$|g_Y M_2 v_\rho| \sim 10^{-5} M_{\rm W}^2$.
\\

Substituting  $v_\rho$ back into the potential adds an extra
contribution to the Higgs potential. The minimization of the
effective potential with respect to the $h_1^0$ and $h_2^0$ gives
two conditions and one combination of these is affected by $\rho$,
and one has
\beqn
\frac{1}{2} M_0^2
&=&\frac{m_1^2-m_2^2\tan^2(\beta)}{\tan^2(\beta) -1} + \frac{g_Y M_2 v_\rho}{2\cos (2\beta)}\ .
\eeqn
where $M_0^2=(g_2^2+g_Y^2)v^2/4$, so that $M_0$ coincides with the Standard Model tree level
prediction in the limit when the \st effects vanish.
In the absence of the \st effect on has $v_\rho =0$ and one recovers
the well known  result of radiative breaking of the
electro-weak symmetry in SUGRA models \cite{sugra}. We see now that the
Stueckelberg effect modifies the equation that determines $M_0^2
\sim M_{\rm Z}^2$, but only by a tiny correction.
\\

Inserting the vacuum expectation values back into the potential,
we compute now the mass matrix for the neutral Higgs fields. The
CP-odd neutral Higgs is not affected by the \st extension.
However, in the CP-even sector one has three states, i.e. $\phi_1
= \Re(h^0_1), \phi_2=\Re(h^0_2)$, and $\rho$, which mix. The
coupling of the $h_i^0$ to $\rho$ adds off-diagonal bilinear
interactions $h_i \rho$. In terms of the basis
$S_a=(\phi_1,\phi_2,\rho)_a^T$ for the CP-even neutral scalars,
the mass term reads
\beqn
-\frac12 \sum_{a,b=1}^3 S_a M^{[0]2}_{ab}S_b
\eeqn
with the following mass matrix
(using the upper index $[0]$ for spin 0)
\beqn
 M^{[0]2}_{ab} =
\left[\matrix{
M_0^2c^2_{\beta} +m_A^2 s^2_{\beta}
   &  -(M_0^2+m_A^2)s_{\beta}c_{\beta}  & -  \frac12 g_Y M_2 v c_{\beta} \cr
-(M_0^2+m_A^2)s_{\beta}c_{\beta}& M_0^2s^2_{\beta} + m_A^2 c^2_{\beta} &
 \frac12 g_Y M_2 v s_{\beta} \cr
- \frac12 g_Y M_2 v c_{\beta} & \frac12 g_Y M_2 v s_{\beta} & M_{\rho}^2}\right]_{ab} \ ,
\label{higgsmasses}
\eeqn
where $(s_{\beta},\, c_{\beta})=(\sin(\beta),\, \cos(\beta))$.
The eigenstates we denote by
\beqn
E^{[0]}_a = (H_1^0, H_2^0, H_3^0)_a^T\ ,
\eeqn
and arrange them so that
\beqn
E^{[0]}_a ~\longrightarrow~ (H^0, h^0, \rho)_a^T\ ,
\eeqn
when $\delta\rightarrow 0$. Then $h^0$ is the light neutral Higgs
of the MSSM and $H^0$ is the heavy one. Instead of two, we now
have three neutral Higgs states, all of which are CP-even states
in resonant production in the $q\bar q$ channel. In CP-violating
channels the number of states will increase to four, since the above
three CP-even states will mix with the CP-odd state $A^0$.
The effect of the mixing on the mass eigenvalues of $h^0$ and $H^0$ 
is governed roughly by the ratios $g_Y M_2 v / m_{i}^2$, for $m_i=M_0,M_{\rho},m_A$, 
and is model-dependent. The correction  on the lightest Higgs 
boson mass could be either positive or negative.
For example, it turns out negative when $\tan(\beta)$ is large 
and $M_{\rho} >M_0$. The size of correction could be as large as 
a few GeV but significantly smaller than the loop corrections. 
A quantitative analysis requires a global fit to the electro-weak 
data and is beyond the scope of the present work.\\


The new state that appears above is $H_3^0$ which has the quantum
numbers $J^{\rm CP}=0^+$. This state is mostly the $\rho$ state
and its decay into visible sector will be dominantly into $t\bar
t$, provided $m_{H^0_3}>2m_t$, or otherwise into $b\bar b$. We
expect the size of the relevant mixing parameter to be ${\cal
O}(M_2/M_1)\sim 0.01$ and thus the decay width will be in the
range of MeV or less. The  production of such a resonance in
$e^+e^-$ colliders will be difficult since the couplings of this
state to fermions is proportional to the mass and in addition
there are suppression factors. A possible production mechanism is
at hadron colliders via the Drell-Yan process using the $q\bar q
H_3$ vertex, where the largest contributions will arise when
$q=(b,t)$.


\subsubsection{Stueckelberg corrections to sfermion masses}

The Stueckelberg  effect modifies the D-term correction to squark
and slepton masses. This can be seen by examining the effective
lagrangian after elimination of $D_B$ and $D_C$. The effective
potential then is
\beqn
{\cal V}(\tilde q_i, \tilde l_i,\rho) =
\frac{1}{2} \left[ \sum_{i} \bar z_i \frac{g_Y Y}{2} z_i \right]^2
+ \rho M_2 \sum_{i} \left( \bar z_i \frac{g_Y Y}{2} z_i \right)\ .
\eeqn
The D-term correction to the mass of the sfermion $z_i$ is
\beqn
\Delta \tilde m_{z_i}^2 =\frac{Y_i}{2} v_\rho g_Y M_2 +
\frac{Y_i}{2}\sin^2(\theta_W) \cos (2\beta) M_0^2
\eeqn
Of course, to the above we must add the D-term correction from
$SU(2)_L$ sector.  Finally, we note that an interesting sum rule
results in the case when $M_2/M_1\ll 1$ relating the $\rho$ mass
and the ${\rm Z}'$ mass. In this limit one finds from
Eq.(\ref{zmasses}) and Eq.(\ref{higgsmasses}) the following
approximate sum rule
\beqn
M_{\rho}^2  ~\simeq~
M_{{\rm Z}'}^2 + m_{\rho}^2
\eeqn
Clearly, $M_{\rho} \geq M_{{\rm Z}'}$, the additional spin zero
state is heavier than the ${\rm Z}'$ boson.


\subsection{The fermionic sector of the extended MSSM}

We discuss now the fermionic sector of the theory. For the neutral
fermions instead of four neutral Majorana fields in the MSSM, we
have  a set of six fields. These consist of the three gauginos,
the two Higgsinos $\tilde h_i$, and the extra Stueckelberg
fermions $\psi_S$. We order the six neutral fields into a vector
$\psi$
\beqn
\psi_a = (\psi_S,\lambda_X,\lambda_Y,\lambda_3,\tilde h_1,\tilde h_2)^T_a
\eeqn
and write the mass term as (upper index $[1/2]$ for spin $1/2$)
\beqn
-\frac{1}{2} \sum_{a,b=1}^6 \bar \psi_a M^{[1/2]}_{ab} \psi_b\ .
\eeqn
From the \st correction  to the MSSM Lagrangian, the only
correction is due to the coupling of $\psi_S$ to gauginos, because
the triliniear coupling with the scalars $z_i$ does not induce
bilinear fermion interactions, as $\langle z_i \rangle=0$. After
spontaneous breaking of the electro-weak symmetry the neutralino
mass matrix in the above basis is given by
\beqn \label{neutrmass}
M^{[1/2]}_{ab} =
\left[\matrix{ 0 & M_1 & M_2 & 0 & 0 & 0\cr
M_1& \tilde m_S & 0 & 0 & 0 & 0\cr
M_2& 0 & \tilde m_1 & 0 & -c_{\beta}s_{W}M_0 & s_{\beta}s_WM_0\cr
0 & 0 & 0 & \tilde m_2 & c_{\beta}c_{W}M_0 & -s_{\beta}c_WM_0 \cr
0 & 0 & -c_{\beta}s_{W}M_0  &  c_{\beta}c_{W}M_0 & 0 & -\mu \cr
0 & 0 & s_{\beta}s_{W}M_0  &  -s_{\beta}c_{W}M_0 &  -\mu & 0}\right]_{ab} \ ,
\eeqn
We note that the zero entry in the  upper left hand corner arises due to the Weyl fermions
not acquiring soft masses.
The above gives rise to six Majorana mass eigenstates which
we label as follows
\beqn
E^{[1/2]}_a = (\chi_1^0,\chi_2^0,\chi_3^0,\chi_4^0,\chi_5^0, \chi_6^0)^T_a
\eeqn
The two additional Majorana eigenstates $\chi_5^0, \chi_6^0$ are
due to the Stueckelberg extension. To get an idea of the effect of
the Stueckelberg sector, we exhibit the eigenvalues in the limit
when $M_{\rm Z}$ is negligible relative to all other mass
parameters in the mass matrix. In this case the spectrum consists
of
\beqn
m_{\chi_i^0} ~(i=1-4),~~
m_{\chi_5^0}=\sqrt{M_1^{2} +\frac{1}{4}\tilde m_S^{2}}
+\frac{1}{2} \tilde m_S,~~
m_{\chi_6^0}=\sqrt{M_1^{2} +\frac{1}{4}\tilde m_S^{2}}
-\frac{1}{2} \tilde m_S
\eeqn
where $m_{\chi_i^0} ~(i=1-4)$ are the four eigenvalues that arise
from diagonalization of the $4\times 4$ mass matrix in the lower
right hand corner. These are the usual eigenvalues that one has in
the MSSM. The eigenvalues $m_{\chi_5^0}$ and $m_{\chi_6^0}$
correspond to the  heavy and light additional states which we
christen as Stueckelberginos. For the case when
$m_{\chi_{5,6}^0}>m_{\chi_1^0}$ not much will change, and the
analysis of dark matter will essentially  remain unchanged.
However, for the case when the light Stueckelbergino is lighter
than the lightest of $m_{\chi_i^0} ~(i=1-4)$, then the situation
is drastically changed. In this case the lightest supersymmetric
particle (LSP) is no longer a neutralino, i.e. of the set
$m_{\chi_i^0} ~(i=1-4)$, but rather the Stueckelbergino $\chi_{\rm
St}^0=\chi_6^0$.
\\

\begin{figure}[h]
\begin{center}
\vbox{ \psfig{figure=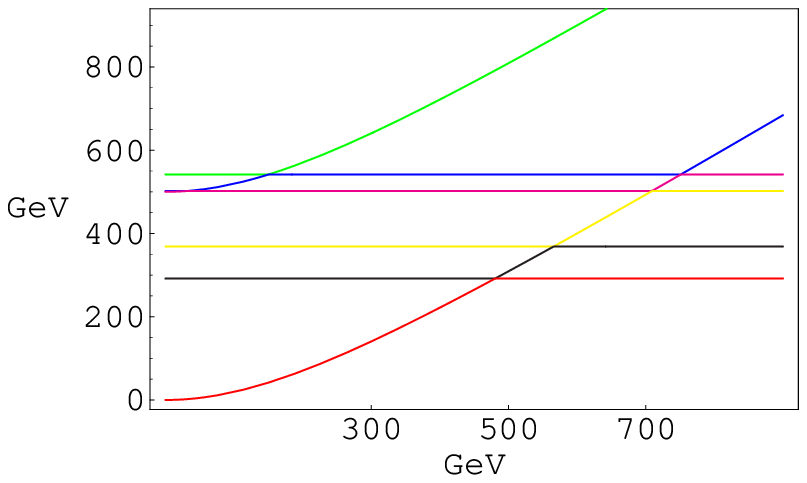,width=7cm}
\psfig{figure=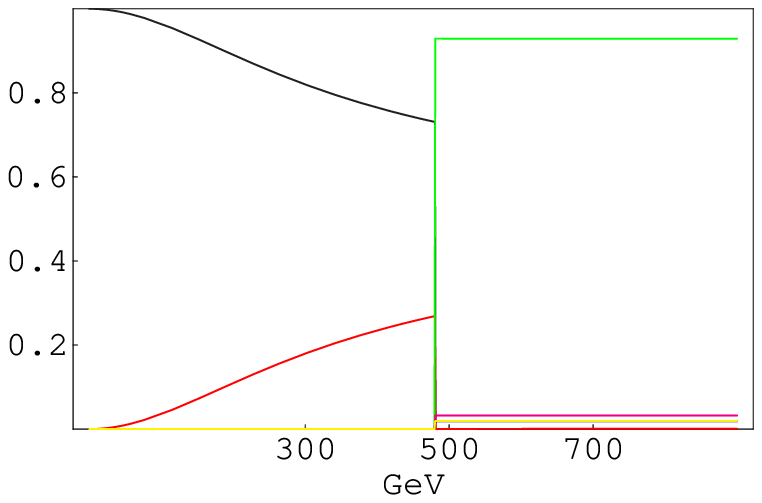,width=7cm} } \caption{Plot of the
neutralino mass spectrum as a function of $M$ (left), for values
$\tan(\beta)=3$, $\mu = 500$, $m_2 = 400$, $m_1 = 300$, $m_S =
500$, $\delta  = 0.029$ and of the (squared) components of the LSP
also as a function of $M$ (right).}\label{spectrum_lsp}
\end{center}
\end{figure}
\begin{figure}[h]
\begin{center}
\resizebox{8cm}{!}{\psfig{figure=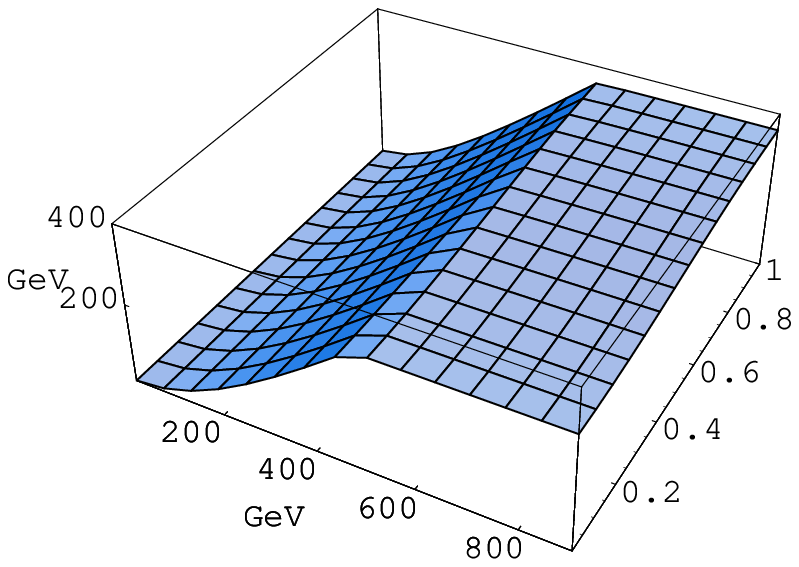,width=8cm}}
\caption{The mass of the LSP as function of $M$ and $\delta$. Plot of the lowest
eigenvalues of the neutralino mass matrix for
values $\tan(\beta)=3$, $\mu = 500$, $m_2 = 400$, $m_1 = 300$,
$m_S = 500$ as a function of $M$ and $\delta$.}\label{3dmass}
\end{center}
\end{figure}

We illustrate the above phenomena in Figure \ref{spectrum_lsp}. In
the left part of this figure the masses of the six neutralinos are
plotted as a function of $M$ for the inputs given there. For
values of $M$ above  around $500\, {\rm  GeV}$ the LSP is the
usual MSSM neutralino and its mass is essentially unaffected by
$M$. However, as we move to values of $M$ below $500\, {\rm GeV}$,
one finds that there is a sudden transition and the LSP becomes
mostly a \st fermion and its mass then varies rapidly with $M$.
The same phenomenon is illustrated in the figure to the right
where the square of the magnitudes of the components of the LSP in
spectral decomposition are plotted, i.e. one writes the LSP
($\chi^0$) as follows
\beqn
\chi^0= C_S \psi_S +C_X \lambda_X + C_Y \lambda_Y + C_3 \lambda_3 + C_1\tilde h_1+C_2\tilde h_2\ .
\eeqn
Thus, on the right hand side  Figure \ref{spectrum_lsp} below
$M=500\, {\rm  GeV}$ the upper curve is $|C_S|^2$ and the lower
curve is $|C_X|^2$ while the other components are too small to be
visible. Above $M=500\, {\rm  GeV}$, the upper curve is $|C_Y|^2$
while the next lower curve is $|C_3|^2$ etc. Again in this figure
we see a rather sudden transition from the LSP being an almost
pure MSSM particle above $M=500\, {\rm  GeV}$ to being an almost
\st fermion below $M=500\, {\rm  GeV}$.  Another view of the same
phenomenon is given in  Figure \ref{3dmass}  in a plot showing the
LSP mass along the vertical axis versus values of $M$ and $\delta$
along the horizontal axes. It displays the very weak dependence of
the mass of the lightest neutralino eigenstate on $\delta$ over
basically the whole range of allowed parameters, while there is a
significant bending at around $M=500\, {\rm  GeV}$ in the
dependence on $M$.
\\

If indeed $\chi_{\rm St}^0$ is  the LSP then aside from the issue of
a re-analysis of dark matter, the supersymmetric signals would be
drastically modified. The usual missing energy signals where the
lightest neutralino $\chi_1^0$ is the LSP do not apply. Indeed if
$\chi_{\rm St}^0$ lies lower than $\chi_1^0$, then $\chi_1^0$  will
be unstable and will decay into $\chi_{\rm St}^0$ by a variety of decays
channels such as
\beqn
\chi_1^0~~\rightarrow~~ q_i\bar q_i
\chi_{\rm St}^0\ , ~~{\it l}_i\bar {\it l}_i \chi_{\rm St}^0\ ,~~
{\rm Z}\chi_{\rm St}^0\ .
\eeqn
We estimate the lifetime for this decay to lie in the range
$10^{-19\pm 2}$ sec. Thus $\chi_{1}^0$ will decay in the detection
chamber. In this case the detection signals will change
drastically. Thus, for example, the decay of the chargino
$\chi_1^{-}\rightarrow {\it l^-}+\{\chi_1^0 + \nu_{\it l}\} $ will
be changed into $\chi_1^{-}\rightarrow {\it l_1^-}{\it l_2^-}{\it
l_2^+} +\{\chi_{\rm St}^0 + \nu_{\it l}\}$.  Similarly the decay
of the slepton will lead also to a possible three lepton final
state, i.e. $\tilde{\it l}^{-} \rightarrow {\it l^-} {\it l_1^-}
{\it l_1^+ } +\{ \chi_{\rm St}^0 + \nu_{\it l} \}$ while the well
known decay of the off-shell W, i.e. ${\rm W}^*\rightarrow
\chi_1^-\chi_2^0$, which in SUGRA models gives a trileptonic
signal \cite{trilep}, in the present context can give rise to
final states with three, five and seven leptons. Thus, we see that
in this case there will be quite a significant change in the
analysis of the phenomenology in search for supersymmetry.
However, if the mass difference between $\chi_1^0$ and $\chi_{\rm
St}^0$ is not substantial, then the $q_i\bar q_i$ and ${\it
l}_i\bar {\it l}_i$ produced in the decay of the $\chi_1^0$ may be
too soft to be detected. In this case there would be no
substantial change in the SUSY signatures.
\\

Also of interest is the status of dark matter in the Stueckelberg
extension. As noted above there are now six neutral fermionic
states compared with four for the case of the MSSM. The parameter
space of the model is now also larger involving in addition to the
MSSM parameters also the parameters of the Stueckelberg sector. It
is known that in mSUGRA model over a significant part of the
parameter space the lightest MSSM neutralino is also the LSP and
thus  a candidate for cold dark matter (CDM). This is also the
case in the Stueckelberg extension. There exists a significant
part of the parameters space where the lightest MSSM neutralino is
the LSP. In this case the lightest MSSM neutralino will still be
the cold dark matter candidate and essentially all of the analysis
on dark matter in supergravity models will go through. However,
there exists a part of the parameter space of the Stueckelberg
extension, where the Stueckelberg fermion can become the LSP as
discussed  above. In this case, the analysis of dark matter will
change drastically. A detailed analysis of the relic density is
outside the scope of the present work and requires a separate
analysis.


\subsection{Coupling of gauge bosons to the hidden sector}

While the couplings of the ${\rm Z}'$ boson to the visible sector quarks and
leptons are suppressed because of small mixing angles, this is
not the case for the couplings of the ${\rm Z}'$ boson to the hidden
sector fields. Thus, for example, the couplings of ${\rm Z}'$ to the hidden
sector current $J_X^{\mu}$ is given by
\beqn
{\cal L}_{{\rm Z}'\cdot {\rm hid}} =
\big[\cos(\psi) \cos(\phi) - \sin(\theta)\sin(\phi) \sin(\psi)\big]\,  {\rm Z}_{\mu}'g_XJ_X^{\mu}
\eeqn
which we can rewrite into chiral components, using
\beqn \label{lrcpl}
g_X J_X^{\mu} =
      \sum_i \Big[ g_L^{i} \bar f_{{\rm hid},i} \gamma^\mu (1-\gamma_5) f_{{\rm hid},i}
        + g_R^{i} \bar f_{{\rm hid},i} \gamma^\mu (1+\gamma_5) f_{{\rm hid},i} \Big]\ .
\eeqn
Using the above the decay width of Z$'$ into hidden sector fermions is given by
\beqn
\Gamma ( {{\rm Z}'} \rightarrow f_{{\rm hid}}\bar f_{{\rm hid}} ) &=&
\\
&&
\hspace{-2cm}
\frac{M_{{\rm Z}'}}{6\pi}
\big[\cos(\psi) \cos(\phi) - \sin(\theta)\sin(\phi) \sin(\psi)\big]^2 \sum_i (
 (  g_L^i )^2 + (  g_R^i )^2 )
\label{hidden}\  .
\nonumber
\eeqn
To get an estimate of the ${\rm Z}'$ decay width into the hidden sector
matter, we set $(g_L^i)^2/4\pi=(g_R^i)^2/4\pi\sim 10^{-2}$ and
$M_{{\rm Z}'}=250$ GeV, which gives $\Gamma ( {{\rm Z}'} \rightarrow
f_{\rm hid}\bar f_{\rm hid}) \lesssim 3\, {\rm GeV}$. This is to be compared with the decay
width of the ${\rm Z}'$ into visible sector quarks which lies in the MeV
range. Thus we see that the decay of the ${\rm Z}'$ into hidden sector
matter is much larger compared to the decay width of the ${\rm Z}'$ into
visible sector matter. This is to be expected due to the fact that
${\rm Z}'$ is dominantly composed of $C_{\mu}$ which couples with normal
strength to the hidden sector matter.
\\

Another implication of the Stueckelberg extension is that it
implies the photon couples with the hidden sector, if such a sector
exists, with a small typically irrational charge. Thus one may
write this coupling in the form
\beqn
 \cl_{A^\gamma\cdot {\rm hid}} ~=~ e' A_{\mu}^{\gamma} J^{\mu}_X
\eeqn
where $e'=-g_X\cos(\theta)\sin(\phi)$. We note that similar mini-charges arise
in models with kinetic energy mixings \cite{holdom},
where there are stringent limits on the size of these charges.
 These limits depend critically on the masses of the mini-charged
particles \cite{moha,goldberg}. Although the mechanism by which
the mini-charges arise in the Stueckelberg model discussed here is
very different, one expects similar constraints on the charges of
such particles. However, as discussed at the end of section 7, the
size of the mini-charge with which the photon couples with the
hidden sector is highly model-dependent. Of course, having  a
hidden sector is optional and one may eliminate it altogether by
setting $J_X^{\mu}=0$.


\subsection{Corrections to $g_{\mu}-2$}

Since the Z interactions are modified in the Stueckelberg
extension, there is a modification of the Z exchange contribution
to $g_{\mu}-2$. Further, there is an additional contribution to
$g_{\mu}-2$ from the ${\rm Z}'$ exchange. We now compute these
corrections. For the Z exchange contribution we find
\beqn
\Delta g_{\mu}^{Z}=\frac{m_{\mu}^2 G_F }{12\pi^2 \sqrt 2}
[(3-4\cos^2\theta_W)^2\beta_v^2-5\beta_a^2]
\label{gz1}\ ,
\eeqn
where
\beqn
\beta_v &=& \frac{\sqrt{g_2^2+g_Y^2}}{-g_2^2+3g_Y^2}
\Bigg[ \frac{-g_2^2+3\cos^2\phi g_Y^2}{\sqrt{g_2^2+\cos\phi g_Y^2}}
\cos\psi - 3 g_Y \sin\phi \sin\psi \Bigg]\ ,
\nonumber\\
\beta_a &=& \Big[ \frac{g_2^2+\cos^2\phi g_Y^2}{g_2^2+g_Y^2}\Big]^{\frac{1}{2}}
\Bigg[\cos\psi - \frac{g_Y}{\sqrt{g_2^2+\cos^2\phi g_Y^2}}  \sin\phi \sin\psi \Bigg]
\label{gz2}\ ,
\eeqn
where $G_F=(g_2^2+g_Y^2)/(4\sqrt 2 M_Z^2)$.
For the ${\rm Z}'$ exchange contribution we find
\beqn
\Delta g_{\mu}^{{\rm Z}'}=\frac{m_{\mu}^2 G_F M_Z^2}{12\pi^2 \sqrt 2 M_{{\rm Z}'}^2}
[(3-4\cos^2\theta_W)^2\gamma_v^2-5\gamma_a^2]
\label{gz3}\ ,
\eeqn
where
\beqn
\gamma_v &=& \frac{\sqrt{g_2^2+g_Y^2}}{-g_2^2+3g_Y^2}
\Bigg[ \frac{-g_2^2+3\cos^2\phi g_Y^2}{\sqrt{g_2^2+\cos\phi g_Y^2}}
\sin\psi + 3 g_Y \sin\phi \cos\psi \Bigg]\ ,
\nonumber\\
\gamma_a &=& \Big[\frac{g_2^2+\cos^2\phi g_Y^2}{g_2^2+g_Y^2}\Big]^{\frac{1}{2}}
\Bigg[\sin\psi + \frac{g_Y}{\sqrt{g_2^2+\cos^2\phi g_Y^2}}  \sin\phi \cos\psi \Bigg]
\label{gz4}\ .
\eeqn
The SM limit is $\phi =0 =\psi$, or $\beta_v=1=\beta_a$, $\gamma_v=0=\gamma_a$,
and  Eq.(\ref{gz1}) gives the well known result of the Z exchange
contribution in the SM.
Numerically, the deviations from the SM are significantly smaller than
 the SM Z contribution and
thus not discernible at the current level of hadronic error
\cite{hagiwara,davier} and experimental accuracy \cite{deng} in
the determination of $g_{\mu}-2$.


\section{Stueckelberg at a Linear Collider}

There is a general consensus that the high energy collider to be
built after the Large Hadron Collider (LHC) should be a Linear
Collider \cite{Danielson:1996tf} and may most likely be an
International Linear Collider (ILC) \cite{acco}. The design
energies of such a machine could be $\sqrt s=500$ GeV (NLC500)
with a luminosity of as much as $50fb^{-1}yr^{-1}$ or even larger.
In addition to being an ideal machine for detailed studies of the
properties of low lying supersymmetric particles such as light
chargino and and light sfermions, a linear collider is also an
ideal machine for testing some features of the type of extension
of the SM and of MSSM discussed here.


\subsection{Cross-sections including the Z$'$ pole}

In the following we investigate such phenomena, the possibility of discovering
the extra ${\rm Z}'$ boson arising in the Stueckelberg extension. We begin by computing the
scattering cross section of the process
\beqn
e^+(p_1)+e^-(p_2)\rightarrow \mu^+(q_1)+\mu^-(q_2)\ .
\eeqn
This process can proceed via the direct channel  exchange of the
photon, of the Z and of the ${\rm Z}'$ boson. Using the Lagrangian
for the Stueckelberg extension, an analysis for the spin averaged
differential cross section gives\footnote{As is conventional we
have used the Breit-Wigner parametrization of the amplitudes near
the Z and ${\rm Z}'$ poles in the form used in the fits of the LEP
and the Tevatron data.}
\beqn
\frac{d\sigma}{d\Omega}(e^+e^-\rightarrow \mu^+\mu^-)
&=& \\
&& \hspace{-3cm}
~~~\frac{\pi \alpha^2}{2s} (1+z^2)
+\frac{\alpha}{2\sqrt 2} \frac{G_FM_Z^2(s-M_Z^2)}{((s-M_Z^2)^2+\za)}
(v_ev_{\mu}(1+z^2)+2a_ea_{\mu}z)
\non
&&\hspace{-3cm}
+ \frac{G_F^2M_Z^4s}{16\pi ((s-M_Z^2)^2+\za)}
((v_e^2 +a_e^2) (v_{\mu}^2+a_{\mu}^2)(1+z^2)+ 8v_ea_e v_{\mu}a_{\mu}  z)
\non
&&\hspace{-3cm}
+\frac{\alpha}{2\sqrt 2} \frac{G_FM_Z^2(s-M_{{\rm Z}'}^2)}{((s-M_{{\rm Z}'}^2)^2+\zb)}
(v_e'v_{\mu}'(1+z^2)+2a_e'a_{\mu}'z)
\non
&&\hspace{-3cm}
+ \frac{G_F^2M_Z^4s}{16\pi ((s-M_{{\rm Z}'}^2)^2+\zb)}
((v_e^{'2} +a_e^{'2}) (v_{\mu}^{'2}+a_{\mu}^{'2})(1+z^2) + 8v_e'a_e'v_{\mu}'a_{\mu}'z) \non
&&\hspace{-3cm}
+ \frac{G_F^2M_Z^4 s(s-M_Z^2)(s-M_{{\rm Z}'}^2)}{8\pi ((s-M_{Z}^2)^2+\za) ((s-M_{{\rm Z}'}^2)^2+\zb) }
\non
&& \hspace{-2.5cm}
\times
( (v_{\mu}v_{\mu}'+a_{\mu}a_{\mu}')(v_ev_e'+a_ea_e') (1+z^2) + 2(v_{\mu}a_{\mu}' +a_{\mu}v_{\mu}')
(v_ea_e'+a_ev_e')z)\ ,
\nonumber
\label{t1}
\eeqn
where $z=\cos(\vartheta)$ with $\vartheta$ the scattering angle in the center of mass, and
 $v_e, v_{\mu}, $ etc are defined as follows
\beqn
v_e&=&v_{\mu}~=~ (\beta_L+\beta_R)\sin^2(\theta_W)-\frac{1}{2}\beta_L\ ,
\nonumber\\
a_e&=&a_{\mu}~=~ (\beta_L-\beta_R)\sin^2(\theta_W)-\frac{1}{2}\beta_L\ ,
\nonumber\\
v_e'&=&v_{\mu}'~=~ (\gamma_L+\gamma_R)\sin^2(\theta_W)-\frac{1}{2}\gamma_L\ ,
\nonumber\\
a_e'&=&a_{\mu}'~=~ (\gamma_L-\gamma_R)\sin^2(\theta_W)-\frac{1}{2}\gamma_L\ ,
\label{t2}
\eeqn
where
\beqn
\beta_R&=& \Bigg[\frac{1+\tan^2(\theta_W)}{1+\tan^2(\theta_W) \cos^2(\phi)}\Bigg]^{\frac{1}{2}} \cos^2(\phi) \cos(\psi)
-\frac{\sqrt{1+\tan^2(\theta_W)}}{\tan(\theta_W)} \sin(\phi) \sin(\psi)\ ,
\non
\beta_L&=& \frac{\sqrt{1+\tan^2(\theta_W)}}{1-\tan^2(\theta_W)}
\Bigg[\frac{1-\tan^2(\theta_W)\cos^2(\phi)}{\sqrt{1+\tan^2(\theta_W)\cos^2(\phi)}} \cos(\psi)
+ \tan(\theta_W)\sin(\phi) \sin(\psi)\Bigg]\ ,
\non
\gamma_R&=& \Bigg[\frac{1+\tan^2(\theta_W)}{1+\tan^2(\theta_W) \cos^2(\phi)}\Bigg]^{\frac{1}{2}} \cos^2(\phi) \sin(\psi)
+\frac{\sqrt{1+\tan^2(\theta_W)}}{\tan(\theta_W)} \sin(\phi) \cos(\psi)\ ,
\nonumber\\
\gamma_L&=& \frac{\sqrt{1+\tan^2(\theta_W)}}{1-\tan^2(\theta_W)}
\Bigg[ \frac{1-\tan^2(\theta_W)\cos^2(\phi)}{\sqrt{1+\tan^2(\theta_W)\cos^2(\phi)}} \sin(\psi)
                                     - \tan(\theta_W)\sin(\phi) \cos(\psi) \Bigg]\ .
\nonumber
\eeqn
Eq.(\ref{t1}) contains six different type of terms. These consist
of three direct channel poles corresponding to the direct $s$
channel exchange of the photon, the Z boson and the ${\rm Z}'$
boson, and three interference terms which consist of the
interference between the photon and the Z boson exchanges, the
interference between the photon and the ${\rm Z}'$ boson
exchanges, and the interference between the Z boson and the ${\rm
Z}'$ boson exchanges. The entire effect of the Stueckelberg
extension are contained in the parameters $\beta_L$, $\beta_R$,
$\gamma_L$, and $\gamma_R$. Here $\beta_L$ and $\beta_R$ give the
modification of the Z exchange interactions due to the
Stueckelberg extension, and of course the ${\rm Z}'$ interactions
arise exclusively from the Stueckelberg extension.  Thus the SM
limit corresponds to $\beta_L=1=\beta_R$, and
$\gamma_L=0=\gamma_R$. We note that the $\Gamma_{{\rm Z}'}$ can
also get contributions from the decay into the hidden sector.
\\

Also of interest is  the
scattering cross section of the process
\beqn
e^+(p_1)+e^-(p_2)\rightarrow  q(q_1)+\bar q(q_2)\ .
\eeqn
Again this process can proceed via the direct channel  exchange of
the photon, of the Z and of the ${\rm Z}'$ boson. Using the
Lagrangian for the Stueckelberg extension, an analysis for the
spin averaged differential cross section gives
\beqn
\frac{d\sigma}{d\Omega}(e^+e^-\rightarrow q\bar q)
&=& \\
&&\hspace{-3cm}
~~~ \frac{3\pi \alpha^2 Q_q^2}{2s} (1+z^2)
-\frac{3\alpha Q_q}{2\sqrt 2} \frac{G_FM_Z^2(s-M_Z^2)}{((s-M_Z^2)^2+\za)} (v_ev_q(1+z^2)+2a_ea_qz)
\non
&&\hspace{-3cm}
+ \frac{3G_F^2M_Z^4s}{16\pi ((s-M_Z^2)^2+\za)}((v_e^2 +a_e^2) (v_q^2+a_q^2)(1+z^2) + 8v_ea_e v_qa_q  z)
\non
&&\hspace{-3cm}
-\frac{3\alpha Q_q}{2\sqrt 2} \frac{G_FM_Z^2(s-M_{{\rm Z}'}^2)}{((s-M_{{\rm Z}'}^2)^2+\zb)}
              (v_e'v_q'(1+z^2)+2a_e'a_q'z)
\non
&&\hspace{-3cm}
+ \frac{3G_F^2M_Z^4s}{16\pi ((s-M_{{\rm Z}'}^2)^2+\zb)}
   ((v_e^{'2} +a_e^{'2}) (v_q^{'2}+a_q^{'2})(1+z^2) + 8v_e'a_e'v_q'a_q'z)
\non
&&\hspace{-3cm}
+ \frac{3G_F^2M_Z^4 s(s-M_Z^2)(s-M_{{\rm Z}'}^2)}{8\pi ((s-M_{Z}^2)^2+\za) ((s-M_{{\rm Z}'}^2)^2+\zb) }
\non
&&\hspace{-2cm}
\times      ( (v_qv_q'+a_qa_q')(v_ev_e'+a_ea_e') (1+z^2) + 2(v_qa_q' +a_qv_q') (v_ea_e'+a_ev_e')z)
\label{q1}
\nonumber\ ,
\eeqn
where 3 is the color factor,  $Q_u=\frac23$, $Q_d=-\frac13$. In the above
$v_q,a_q$ are defined as follows
\beqn
v_q&=& \frac{1}{2}[\delta_L \tau_3 -2\sin^2(\theta_W) Q_q(\delta_L +\delta_R)]\ ,
\nonumber\\
a_q&=& \frac{1}{2}[\delta_L \tau_3 -2\sin^2(\theta_W) Q_q(\delta_L -\delta_R)]\ ,
\eeqn
where $\tau_3 =(1,-1)$ for $q=(u,d)$ and
\beqn
\delta_R&=&\delta_{\rm em} ~=~ \frac{\sqrt{1+\tan^2(\theta_W) \cos^2(\phi)}}{\sin^2(\theta_W)\sqrt{1+\tan^2(\theta_W)}}
(\sin^2(\theta_W)\cos(\psi)-\sin(\theta)\tan(\phi)\sin(\psi))\ ,\nonumber\\
\delta_L&=& \frac{\delta_3\tau_3-2\delta_{\rm em}Q_q\sin^2(\theta_W)}{\tau_3-2Q_q\sin^2(\theta_W)}\ ,
\non
\delta_3&=& \frac{\sqrt{1+\tan^2(\theta_W) \cos^2(\phi)}}{\sqrt{1+\tan^2(\theta_W)}}
(\cos(\psi)-\sin(\theta)\tan(\phi)\sin(\psi))\ .
\label{q3}
\eeqn
Similarly, in the above $v_q',a_q'$ are defined as follows
\beqn
v_q'&=& \frac{1}{2}[\epsilon_L \tau_3 -2\sin^2(\theta_W) Q_q(\epsilon_L +\epsilon_R)],
\nonumber\\
a_q'&=& \frac{1}{2}[\epsilon_L \tau_3 -2\sin^2(\theta_W) Q_q(\epsilon_L -\epsilon_R)],
\eeqn
where
\beqn
\epsilon_R&=&\epsilon_{\rm em} ~=~ \frac{\sqrt{1+\tan^2(\theta_W) \cos^2(\phi)}}{\sin^2(\theta_W)\sqrt{1+\tan^2(\theta_W)}}
( \sin^2(\theta)\sin(\psi) +\cos(\psi)\sin(\theta) \tan(\phi) )\ ,
\nonumber\\
\epsilon_L&=& \frac{\epsilon_3\tau_3-2\epsilon_{em}Q_q\sin^2(\theta_W)}{\tau_3-2Q_q\sin^2(\theta_W)}\ ,
\non
\epsilon_3&=& \frac{\sqrt{1+\tan^2(\theta_W) \cos^2(\phi)}}{\sqrt{1+\tan^2(\theta_W)}}
(\sin(\psi)+\cos(\psi) \sin(\theta)\tan(\phi))\ .
\eeqn
The SM limit is $\phi=0=\psi, ~\theta=\theta_W$, or
$\delta_3=1, \delta_{\rm em}=1, \epsilon_3=0, \epsilon_{\rm em}=0$, and
\beqn
v_q&=&\frac{1}{2}(\tau_3-4Q_q\sin^2(\theta_W))\ ,
\nonumber\\
a_q&=&\frac{1}{2}\tau_3\ ,
\nonumber\\
 v_q'&=&a_q'~=~ 0\ ,
\eeqn
which is correctly the SM result.


\subsection{Forward - backward asymmetry near the Z$'$ pole}

The forward-backward asymmetry is a  useful tool in identifying  the nature of the
underlying interaction.
One defines it as
\beqn
A_{fb} = \frac{\int_0^1 dz \frac{d\sigma}{dz} -\int_{-1}^0 dz\frac{d\sigma}{dz}}
{\int_{-1}^1 dz \frac{d\sigma}{dz}}\ .
\label{fb}
\eeqn
Consider the case of $e^+e^-\rightarrow \mu^+\mu^-$ scattering.
Here $\sigma_{\mu^+\mu^-} = \int_{-1}^1dz\frac{d\sigma}{dz}$ is
given by
\beqn
 \sigma_{\mu^+\mu^-}
&=&\frac{4\pi \alpha^2}{3s}
+\frac{2\sqrt 2\alpha}{3} \frac{G_FM_Z^2(s-M_Z^2) v_ev_{\mu} }  {((s-M_Z^2)^2+\za)}
+ \frac{G_F^2M_Z^4s  (v_e^2 +a_e^2) (v_{\mu}^2+a_{\mu}^2)} {6\pi ((s-M_Z^2)^2+\za) }
\nonumber\\
&&+\frac{2\sqrt 2\alpha}{3} \frac{G_FM_Z^2(s-M_{{\rm Z}'}^2) v_e'v_{\mu}' }{((s-M_{{\rm Z}'}^2)^2+\zb)}
+ \frac{G_F^2M_Z^4s   (v_e^{'2} +a_e^{'2}) (v_{\mu}^{'2}+a_{\mu}^{'2})}{6\pi ((s-M_{{\rm Z}'}^2)^2+\zb)}
\nonumber\\
&&
+ \frac{G_F^2M_Z^4 s(s-M_Z^2)(s-M_{{\rm Z}'}^2)
 (v_{\mu}v_{\mu}'+a_{\mu}a_{\mu}')(v_ev_e'+a_ea_e') }{3\pi ((s-M_{Z}^2)^2+\za)
((s-M_{{\rm Z}'}^2)^2+\zb) }\ .
\label{sigmamumu}
\eeqn
Using the above we  can write the forward-backward asymmetry for this case so that
\beqn
\sigma_{\mu^+\mu^-} A_{fb}^{\mu^+\mu^-} &=&
\frac{\alpha}{\sqrt 2} \frac{G_FM_Z^2(s-M_Z^2)  a_ea_{\mu}}  {((s-M_Z^2)^2+\za)}
+ \frac{G_F^2M_Z^4s  v_ea_e v_{\mu}a_{\mu}} {2\pi ((s-M_Z^2)^2+\za)}\non
&&
+\frac{\alpha}{\sqrt 2} \frac{G_FM_Z^2(s-M_{{\rm Z}'}^2) a_e'a_{\mu}' }   {((s-M_{{\rm Z}'}^2)^2+\zb)}
+ \frac{G_F^2M_Z^4s v_e'a_e'v_{\mu}'a_{\mu}'   }{2\pi ((s-M_{{\rm Z}'}^2)^2+\zb)}
\nonumber\\
&&
+ \frac{G_F^2M_Z^4 s(s-M_Z^2)(s-M_{{\rm Z}'}^2) (v_{\mu}a_{\mu}' +a_{\mu}v_{\mu}')   (v_ea_e'+a_ev_e')  }
{4\pi ((s-M_{Z}^2)^2+\za) ((s-M_{{\rm Z}'}^2)^2+\zb) }\ .
\eeqn
We now start to discuss the numerical results for the total cross
section at the Z$'$ pole, including or excluding the possibility
of hidden sector matter fields it couples to. At the same time, we
display the modifications of the forward-backward asymmetry near
the pole.
\\

\begin{figure}[h]
\begin{center}
\vbox{ \psfig{figure=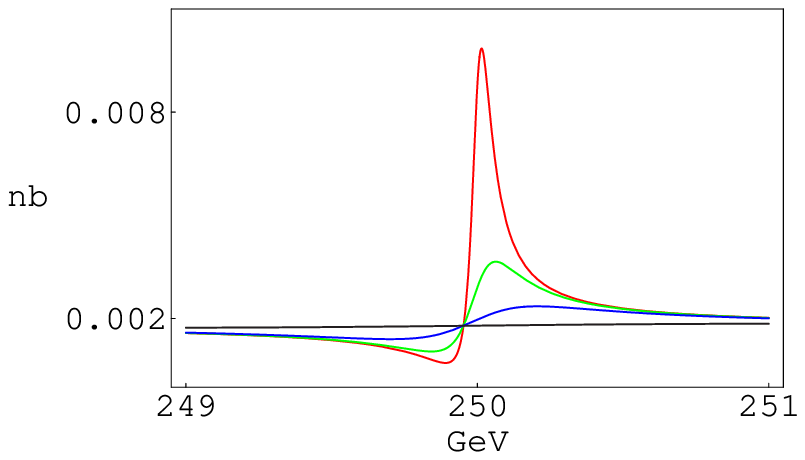,width=7cm}
\psfig{figure=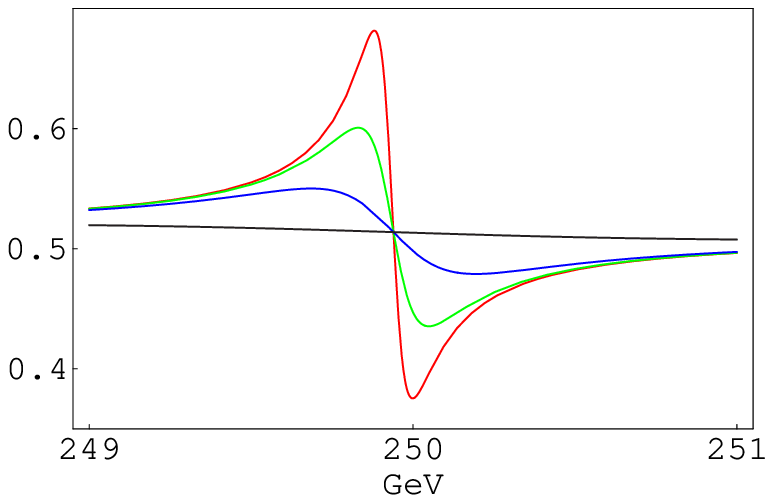,width=7cm} } \caption{Plot of the total
cross-section $\sigma(e^+e^-\rightarrow \mu^+\mu^-)$ (left) and
the forward-backward asymmetry $A_{fb}$ in $e^+e^-\rightarrow
\mu^+\mu^-$ (right) in the vicinity of the Z$'$ resonance for
$M_{{\rm Z}'}=250\,{\rm GeV}$, $\phi= 0.029$. The values of
$\Gamma_{{\rm Z}'}$ are $3\, {\rm GeV}$ (black line), $0.5\, {\rm
GeV}$ (blue line), $0.2\, {\rm GeV}$ (green line), $0.08\, {\rm
GeV}$ (red line).} \label{mupm_sigma_afb}
\end{center}
\end{figure}
\begin{figure}[h]
\begin{center}
\vbox{
\psfig{figure=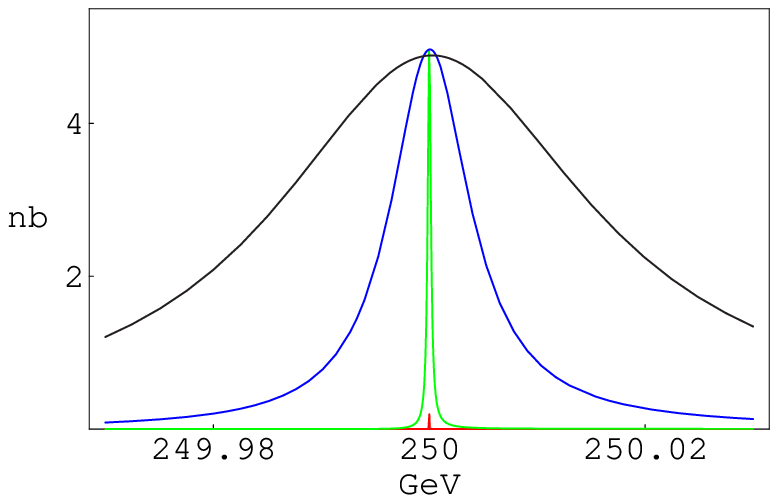,width=7cm}
\psfig{figure=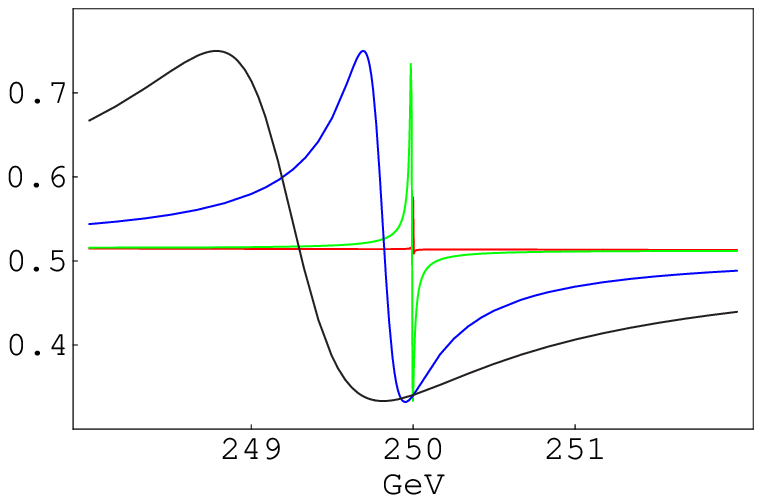,width=7cm}
}
\caption{Plot of the total cross-section
$\sigma(e^+e^-\rightarrow \mu^+\mu^-)$ (left) and
the forward-backward asymmetry $A_{fb}$ in $e^+e^-\rightarrow \mu^+\mu^-$
(right) in the vicinity of the Z$'$ resonance for
$M_{{\rm Z}'}=250\,{\rm GeV}$.
The values of $\delta$ are $0.1$ (black line),
$0.05$ (blue line), $0.01$ (green line),
$0.001$ (red line).}\label{mupm_delta_sigma_afb}
\end{center}
\end{figure}

In Figure \ref{mupm_sigma_afb} we  give a plot of  the
cross-section and $A_{fb}$ for $e^+e^-\rightarrow \mu^+\mu^-$.
First the largeness of the $A_{fb}$ for the SM in this region
comes from the $\gamma -$Z interference term which is large
because of the the axial-vector coupling of the Z boson to
fermions. When the ${\rm Z}'$ contribution is included one finds a
vary rapid variation in the vicinity of the ${\rm Z}'$ pole,
arising from two sources: the ${\rm Z}'$ pole contribution to the
asymmetry and the $\gamma -{\rm Z}'$ contribution to the
asymmetry. These contributions become large in the vicinity of the
${\rm Z}'$ pole and compete in size with the SM contribution from
$\gamma -$Z and Z. In Figure \ref{mupm_sigma_afb} the largest peak
corresponds to the case when there is no hidden sector. The width
$\Gamma_{{\rm Z}'}$ is determined for the decay of ${\rm Z}'$ only
into the fields of the visible sector, and approximating it by
inclusion of only the quark and lepton final states excluding the
top quark contribution. In this case we find that $A_{fb}$ changes
rapidly as we move across the ${\rm Z}'$ pole. Thus, an accurate
dedicated measurement of $A_{fb}$ should give a signal for this
type of resonance.
\\

One may also include a hidden sector in the analysis. This is
easily done by using Eq.(\ref{hidden}). As indicated in the
analysis following Eq.(\ref{hidden}), ${\rm Z}'$ couples  with
normal strength with the fields in the hidden sector and thus the
decay width of the ${\rm Z}'$ into the hidden sector fields need
not be small, and indeed estimates show it to be of size
$\co$(GeV). In Figure \ref{mupm_sigma_afb} we simulate the effects
of the hidden sector by assuming a set of values for $\Gamma_{{\rm
Z}'}$ lying in the range $0.08$ GeV to $3\, {\rm GeV}$. One finds,
as expected, that the effect of including the hidden sector is to
make the peak in $A_{fb}$ near the ${\rm Z}'$ resonance less
sharp. Thus the characteristics of $A_{fb}$ in the vicinity of the
${\rm Z}'$ resonance do indeed carry information regarding the
presence or absence of a hidden sector. In Figure
\ref{mupm_sigma_afb} we also  give a plot of
$\sigma(e^+e^-\rightarrow \mu^+\mu^-)$  in the vicinity of the
${\rm Z}'$ pole. One finds that the cross-section can be much
larger relative to the SM result  near the ${\rm Z}'$ pole. In
Figure \ref{mupm_delta_sigma_afb} an analysis of $A_{fb}$ and of
$\sigma(e^+e^-\rightarrow \mu^+\mu^-)$ for various values of
$\delta$ but without a hidden sector is given. As expected one
finds that the shape of the curves is a very sensitive function of
$\delta$ with the resonance becoming broader as $\delta$
increases. One interesting feature of Figure
\ref{mupm_delta_sigma_afb} is that the peak value of
$\sigma(e^+e^-\rightarrow \mu^+\mu^-)$ is independent of $\delta$.
This is so because the peak value is essentially geometrical in
nature and independent of $\delta$ as long as $\delta$ is small.
This can be easily seen from Eq.(\ref{resonance}) by setting
$E=M_{{\rm Z}'}$. In this limit one finds that ratios $\Gamma({\rm
Z}'\rightarrow e^+e^- (\mu^+\mu^-))/\Gamma({\rm Z}'\rightarrow
all)$ appear. For small values of $\delta$ these ratios are
independent of $\delta$ and take on the value
\beqn
\sigma_{\mu^+\mu^-} (M_{{\rm Z}'}) ~\simeq~
\frac{12\pi}{M_{{\rm Z}'}^2}
\times \left\{ {(\frac{15}{103})^2 \atop (\frac{1}{8})^2} \right.
~=~
\left\{ 4.8\, {\rm nb}~~ {\rm for}~~ M_{{\rm Z}'}<2m_t \atop 3.6\, {\rm nb}~~ {\rm for}~~ M_{{\rm Z}'}>2m_t \right. \ .
\label{peak}
\eeqn
One finds that the analysis of Figure \ref{mupm_delta_sigma_afb}
is consistent with the analytic results on the peak value
corresponding to the case $M_{{\rm Z}'}<2m_t$. We further note
that the drop-off in $\sigma_{\mu^+\mu^-}$ away from the peak is
very sharply dependent on $\delta$. Further, $A_{fb}$ deviates
significantly from the SM prediction over a reasonable domain of
the energy interval and provides another signature for the
discovery of the ${\rm Z}'$ resonance.
\\

\begin{figure}[h]
\begin{center}
\vbox{ \psfig{figure=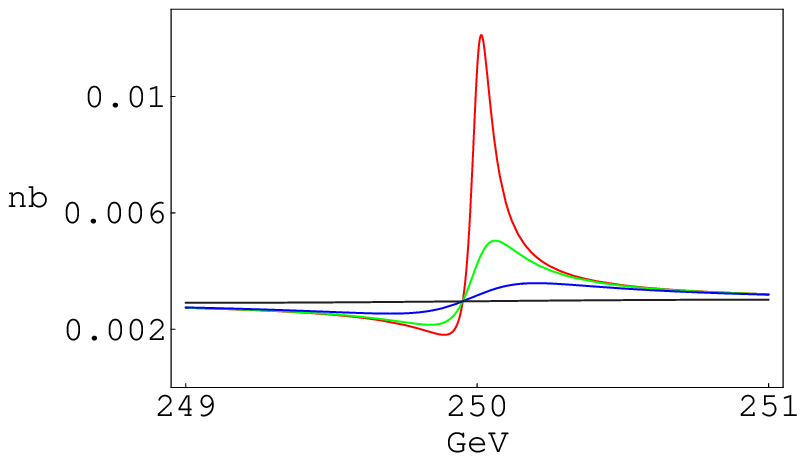,width=7cm}
\psfig{figure=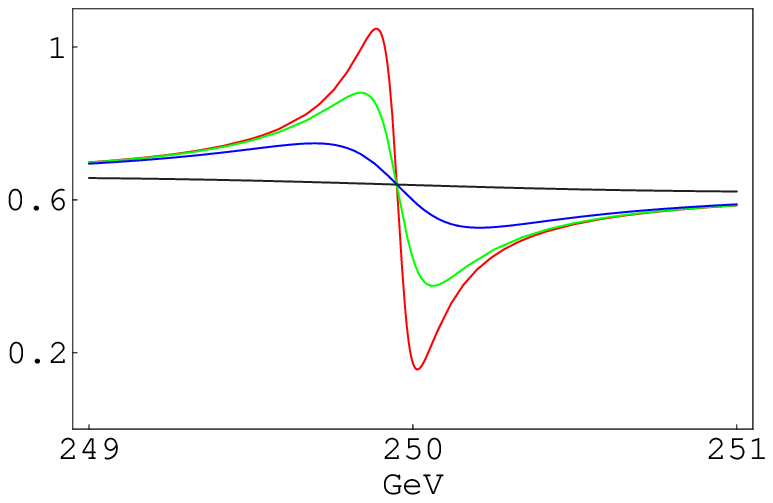,width=7cm} }
\caption{Plot of the total
cross-section $\sigma(e^+e^-\rightarrow u\bar u)$ (left) and the
forward-backward asymmetry $A_{fb}$ in $e^+e^-\rightarrow
u\bar u$ (right) in the vicinity of the Z$'$ resonance for
$M_{{\rm Z}'}=250\,{\rm GeV}$, $\phi= 0.029$. The values of
$\Gamma_{{\rm Z}'}$ are $3\, {\rm GeV}$ (black line), $0.5\, {\rm
GeV}$ (blue line), $0.2\, {\rm GeV}$ (green line), $0.08\, {\rm
GeV}$ (red line).}
\label{uu_sigma_afb}
\end{center}
\end{figure}

A similar analysis can be carried out  for $e^+e^-\rightarrow
q\bar q$. Here we have for the forward-backward asymmetry
\beqn
\sigma_{q\bar q} A_{fb}^{q\bar q} &=&
-\frac{3\alpha Q_q}{\sqrt 2} \frac{G_FM_Z^2(s-M_Z^2)  a_ea_{q}}  {((s-M_Z^2)^2+\za)}
+ \frac{3G_F^2M_Z^4s  v_ea_e v_{q}a_{q}} {2\pi ((s-M_Z^2)^2+\za)}\nonumber\\
&&
-\frac{3\alpha Q_q}{\sqrt 2} \frac{G_FM_Z^2(s-M_{{\rm Z}'}^2) a_e'a_{q}' }   {((s-M_{{\rm Z}'}^2)^2+\zb)}
+ \frac{3G_F^2M_Z^4s v_e'a_e'v_{q}'a_{q}' }{2\pi ((s-M_{{\rm Z}'}^2)^2+\zb)}
\nonumber\\
&&
+ \frac{3G_F^2M_Z^4 s(s-M_Z^2)(s-M_{{\rm Z}'}^2) (v_{q}a_{q}' +a_{q}v_{q}')   (v_ea_e'+a_ev_e')  }
{4\pi ((s-M_{Z}^2)^2+\za) ((s-M_{{\rm Z}'}^2)^2+\zb) }\ ,
\eeqn
where
\beqn
 \sigma_{q\bar q}
&=&\frac{4\pi \alpha^2 Q_q^2}{s}
-{2\sqrt 2\alpha Q_q} \frac{G_FM_Z^2(s-M_Z^2) v_ev_{q} }  {((s-M_Z^2)^2+\za)}
+ \frac{G_F^2M_Z^4s  (v_e^2 +a_e^2) (v_{q}^2+a_{q}^2)} {2\pi ((s-M_Z^2)^2+\za) }
\nonumber\\
&&
-{2\sqrt 2\alpha Q_q} \frac{G_FM_Z^2(s-M_{{\rm Z}'}^2) v_e'v_{q}' }{((s-M_{{\rm Z}'}^2)^2+\zb)}
+ \frac{G_F^2M_Z^4s   (v_e^{'2} +a_e^{'2}) (v_{q}^{'2}+a_{q}^{'2})}{2\pi ((s-M_{{\rm Z}'}^2)^2+\zb)}
\nonumber\\
&&
+ \frac{G_F^2M_Z^4 s(s-M_Z^2)(s-M_{{\rm Z}'}^2) (v_{q}v_{q}'+a_{q}a_{q}')(v_ev_e'+a_ea_e') }{\pi ((s-M_{Z}^2)^2+\za)
((s-M_{{\rm Z}'}^2)^2+\zb) }\ .
\eeqn
A numerical analysis of $A_{fb}$ for the case when the final
states are $u\bar u$ is given in Figure \ref{uu_sigma_afb}, and
again one finds that the characteristics of $A_{fb}$ in the
vicinity of the ${\rm Z}'$ resonance are different for the cases:
$i)$ the SM, $ii)$ the model with a ${\rm Z}'$ resonance but
without hidden sector matter, $iii)$ models including decays of
${\rm Z}'$ into the hidden sector.
\\

\begin{figure}[h]
\begin{center}
\vbox{
\psfig{figure=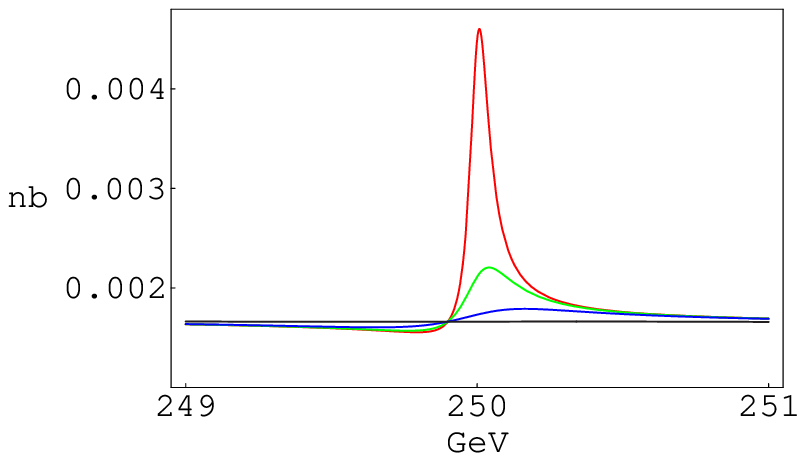,width=7cm}
\psfig{figure=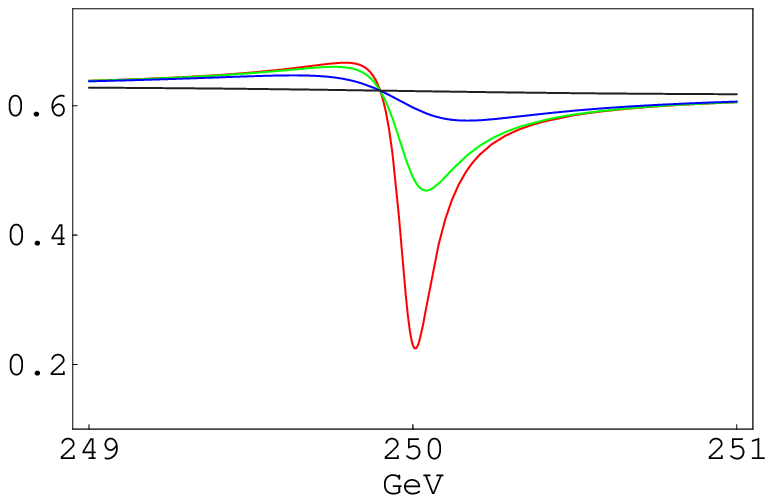,width=7cm}
}
\caption{Plot of the total cross-section
$\sigma(e^+e^-\rightarrow d\bar d)$ (left) and
the forward-backward asymmetry $A_{fb}$ in $e^+e^-\rightarrow d\bar d$
(right) in the vicinity of the Z$'$ resonance for
$M_{{\rm Z}'}=250\,{\rm GeV}$, $\phi= 0.029$.
The values of $\Gamma_{{\rm Z}'}$ are $3\, {\rm GeV}$ (black line),
$0.5\, {\rm GeV}$ (blue line), $0.2\, {\rm GeV}$ (green line),
$0.08\, {\rm GeV}$ (red line).}\label{dd_sigma_afb}
\end{center}
\end{figure}
An analysis of $\sigma(e^+e^-\rightarrow u\bar u)$ is also given
in Figure \ref{uu_sigma_afb}. Here, again  one finds that the
cross section near the vicinity of the pole is significantly
higher than the SM result and the deviation depends on the
presence or absence of the possibility of decays into the hidden
sector. Finally, we discuss the $e^+e^-\rightarrow d\bar d$. In
Figure \ref{dd_sigma_afb} we give  a plot of $A_{fb}$ and one
finds once again very significant deviations from the SM. As in
previous cases the size of the deviation depends on the presence
or absence of ${\rm Z}'$ decays into the hidden sector.
\\

The number of events for the various channels
can be estimated by noting that at the projected design characteristics
of 500 GeV collider one expects an integrated luminosity of $500\, {\rm fb}^{-1}{\rm yr}^{-1}$,
and the number of events using the cross sections of Figures \ref{mupm_sigma_afb}, \ref{uu_sigma_afb}, and
\ref{dd_sigma_afb} are clearly sizable. Finally, we note that
an indication of the presence of a hidden sector to which the ${\rm Z}'$
can decay will be provided by the visible width versus the total width
of the ${\rm Z}'$.


\section{Detection of a sharp ${\rm Z}'$ resonance}

As mentioned already the ${\rm Z}'$  is  expected to be a sharp
resonance, and determination of the $\Gamma({\rm Z}'\rightarrow
e^+e^-)$ is a difficult problem as is well known from the analysis
of the $J/\Psi$ resonance \cite{goldhaber}. A technique which was
useful in the determination of the width of the $J/\Psi$ should
also be valid here, and this is the technique of integrating the
cross section over the resonance \cite{goldhaber}. Thus, for
example, consider the cross-section for the process
$e^+e^-\rightarrow f\bar f$ in the vicinity of the resonance. In
this region one can write the cross-section so that\footnote{Here
we use the simplified form of the Breit-Wigner parametrization.
Use of the more sophisticated form as in Eq.(\ref{t1}) will give
corrections to Eq.(\ref{a1}) only of size $\co(\Gamma_{{\rm
Z}'}^2/M_{{\rm Z}'}^2)$ which are very small.}
\beqn \sigma(e^+e^-\rightarrow f\bar f)=\frac{3\pi}{M_{{\rm Z}'}^2}
\frac{ \Gamma({\rm Z}'\rightarrow e^+e^-)\Gamma({\rm Z}'\rightarrow f\bar f)}
{((E-M_{{\rm Z}'})^2+\frac{1}{4} \Gamma^2({\rm Z}'\rightarrow {\rm all}))} \ ,
\label{resonance}
\eeqn
where
$E=\sqrt s$. Integration over the resonance gives
\beqn
\int
dE\ \sigma(e^+e^-\rightarrow f\bar f)=\frac{6\pi^2}{M_{{\rm Z}'}^2}
 \Gamma({\rm Z}'\rightarrow e^+e^-) \frac{\Gamma({\rm Z}'\rightarrow f\bar f)}
{\Gamma({\rm Z}'\rightarrow {\rm all})}\ .
\label{a1}
\eeqn
For a given final state we define
\beqn
\ca_{\rm fin} =\int dE\ \sigma(e^+e^-\rightarrow {\rm fin})=\frac{6\pi^2}{M_{{\rm Z}'}^2}
\Gamma({\rm Z}'\rightarrow e^+e^-) \frac{\Gamma({\rm Z}'\rightarrow {\rm fin})}{\Gamma({\rm Z}'\rightarrow
{\rm all})}\ ,
\label{a2}
\eeqn
and $\ca_{\rm vis}$ for the sum over all visible final states.
Now for the case of \st ${\rm Z}'$ we have
\beqn
\Gamma({\rm Z}'\rightarrow e^+e^-) ~\simeq~ \frac{\alpha_1}{8}M_{{\rm Z}'}\tan^2(\phi)\ ,
\label{a3}
\eeqn
where $\alpha_1=g_1^2/4\pi$, and we have used the relation
$g_Y'\simeq g_Y=\sqrt{\frac35} g_1$. Further, under the assumption
there is no hidden sector one has\footnote{In the computation of
the ratio in Eq.(\ref{ratios}) we have included only quark and
lepton final states. Inclusion of additional states, specifically
the sparticle final states if they are allowed, will modify these
ratios.}
\beqn \frac{\Gamma({\rm Z}'\rightarrow {\rm vis})}{\Gamma({\rm Z}'\rightarrow {\rm all})}
= \left\{ {\frac{94}{103}~~ {\rm for}~~ M_{{\rm
Z}'}<2m_t \atop \frac{111}{120}~~ {\rm for}~~ M_{{\rm Z}'}>2m_t} \right. \ .
\label{ratios}
\eeqn
Let us now  focus on the final state
$\mu^+\mu^-$. Using Eqs.(\ref{a1}) and (\ref{a3}) we find
\beqn
\ca_{\mu^+\mu^-}^{\rm St} ~=~ \frac{45\pi^2}{412} \frac{\alpha_1\tan^2(\phi)}{M_{{\rm Z}'}}
\label{a5}
\eeqn
For $M_{{\rm Z}'}=250$ GeV, and $\delta=0.02$ one finds
$\ca_{\mu^+\mu^-}^{\rm St} \simeq 1.1 \times 10^{-2}\, $nb-GeV.
Further, we note that the integral of Eq.(\ref{a5}) falls as
$1/M_{{\rm Z}'}$ as $M_{{\rm Z}'}$ gets large. Discovery of the
${\rm Z}'$ depends on the signal versus the background. In this
case the background is the SM contribution. Using the analysis of
Eq.(\ref{sigmamumu}) and excluding the ${\rm Z}'$ contribution one
finds that at $\sqrt s=M_{{\rm Z}'}$,
$\sigma_{\mu^+\mu^-}=1.8\times 10^{-3}\, $nb. If data is collected
in bins of size $\Delta$ (in GeV) then the Standard Model
$\mu^+\mu^-$ cross-section integrated over $\Delta$ around
$M_{{\rm Z}'}$ gives $\ca_{\mu^+\mu^-}^{\rm SM}(\Delta)= 1.8\times
10^{-3} \Delta\, $nb-GeV. Now for larger values of $M_{{\rm Z}'}$
the SM cross section falls as $1/M_{{\rm Z}'}^2$. Putting these
factors together the ratio of the \st contribution to the SM is
given by
\beqn
\frac{\ca_{\mu^+\mu^-}^{\rm St}}{\ca_{\mu^+\mu^-}^{\rm SM}(\Delta)} ~\simeq~ \frac{6}{\Delta ({\rm GeV})}
\Big[\frac{\delta}{0.02}\Big]^2 \frac{M_{{\rm Z}'}({\rm GeV})}{250}\ .
\label{stsm}
\eeqn
The above implies that for $M_{{\rm Z}'}=250$ GeV, and $\delta =0.02$,
the \st effects will give significant enhancement over the
SM result with bin sizes ranging from $1-20\, $GeV.
Further, the signal to background ratio will increase as $M_{{\rm Z}'}$
increases. For example, for $M_{{\rm Z}'}=1\, $TeV, there will be a
further enhancement of roughly a factor of 4. The above
characteristics are encouraging. Of course, the detection of such
an effect will depend on the design characteristics of the machine
such as the beam spread, and the design luminosity.
\\

The result of Eq.(\ref{stsm}) is also encouraging for the search
for a \st ${\rm Z}'$ boson at the Tevatron and at the LHC. Here
one would look for dilepton events in the final state via the
Drell-Yan process and at the Tevatron it is the $e^+e^-$ channel
which would be the most efficient for detection.\footnote{We thank
Darien Wood for pointing this out to us and also for bringing Ref.
\cite{abbot98} to  our attention.} The cross-sections for the
processes $u\bar u\rightarrow l^+l^-$ and $d\bar d\rightarrow
l^+l^-$ given here can be utilized for the computation of the
Drell-Yan production of $e^+e^-$ via the \st ${\rm Z}'$. Our
analysis for the linear colliders hints that the detection of a
sharp ${\rm Z}'$ should also be possible at the hadron colliders.
For example, at the Tevatron the energy resolution is given
roughly by \cite{abbot98} $(15\%/\sqrt{E({\rm GeV})}+1\%)$ where
$E$ for our case would effectively be the di-muon invariant mass.
Thus, for example, for $E=250$ GeV one has a resolution of about
$5\, $GeV. This resolution should allow for a search for a
resonance with characteristics of the type of Eq.(\ref{stsm}).
\\

The radiative return technique might be a useful device to look
for the \st ${\rm Z}'$ resonance. This is a useful procedure when the
colliding beam energies have been fixed to a preassigned value and
not continuously adjustable. In this case one uses initial state
radiation (ISR) to reduce the effective center-of-mass
energy \cite{rr}. Thus consider the process $ e^+e^-\rightarrow
\gamma +{\rm hadrons}$, where the $\gamma$ is a hard photon which
is emitted by one of the initial particles, and is responsible for
reducing the center-of-mass energy. The method allows one to
investigate the entire energy region below the highest energy down
to the threshold. However, appropriate corrections must be made to
account for the possibility that the photon may be emitted by the
final state, i.e. one must take into account the final state
radiation (FSR). One advantage of this technique is that the
systematics of measurements remain unchanged in the scan as one
changes the energy while in conventional energy scans systematics
must be fixed at each step.


\section{Stueckelberg extension with many extra $U(1)$}
\label{sec7}

The Stueckelberg technique is, of course, extendable to more than
one extra $U(1)$ gauge symmetry. In orientifold string compactifications with
D-branes the number of axions in the model is derived from the
dimensional reduction of the ten-dimensional RR forms in the
spectrum of the theory, and given by some topological quantity,
the number of relevant homological cycles of the internal space.
In principle it is an arbitrary number.\footnote{This is 
actually similar for the heterotic string, which was recently
demonstrated in \cite{Blumenhagen:2005ga}.} For example, in
the so-called intersecting D-brane models on toroidal backgrounds,
it was found that four such scalars participate in the generalized
Green-Schwarz mechanism, and may thus also couple to the abelian
gauge fields of the model in form of the \st Lagrangian. In
general, one may write the extended Lagrangian with $N_V$ abelian
gauge fields and $N_S$ axions
\beqn
{\cal L}_{\rm St} = -\frac{1}{4} \sum_{i=1}^{N_V} \Big( C_{\mu\nu i}C_i^{\mu\nu}+ g_i C_{\mu i} J^\mu_i\Big)
 -\frac12 \sum_{j=1}^{N_S} \Big( \partial_\mu \sigma_j + \sum_{i=1}^{N_V} M_{ij} C_{\mu i} \Big)^2
\ .
\eeqn
We have now summarized the hyper charge gauge boson as one among the abelian gauge fields,
say for $i=1$ we let $B_\mu = C_{\mu 1}$.
The generalized $U(1)^{N_V}$ gauge invariance is given by
\beqn
\delta_i C_{\mu i} = \partial_\mu  \lambda_i \ , \quad
\delta_i \sigma_j = - M_{ij} \lambda_i \ .
\eeqn
In a very similar vein one can extend the supersymmetric minimal model by many axions and many
abelian gauge bosons, as in
\beqn
{\cal L}_{\rm St} = \int d^2\theta d^2\bar \theta \,
\sum_{j=1}^{N_S} \left( S_j +\bar S_j + \sum_{i=1}^{N_V} M_{ij} C_i \right)^2 \ ,
\eeqn
where $S_j$ and $C_i$ are the chiral and vector multiplets that
include the axions and gauge fields. One can now easily see that
the effect of each axion is to give mass to exactly one gauge
boson, at least generically. The mass term induced after gauge
fixing is a sum of squares, and each linear combination of masses
$M_{ij}$, reading the $N_S\times N_V$ matrix as a set of vectors in the
$N_V$-dimensional space of abelian gauge fields, defines one massive
direction. In other words, the kernel of $M_{ij}$ defines the set
of the surviving massless abelian vectors. So, generically, all
axions will be eaten by vectors, and only if there are more
vectors than axions ($N_V>N_S$), or linear relations among their
couplings will there be abelian gauge symmetries
surviving.\footnote{This is an important constraint in the
construction of string theoretic brane world models, where one has
to impose extra constraints on the brane configurations to achieve
this.} If we further add the spontaneous electro-weak symmetry
breaking through the Higgs mechanism, there is one more degree of
freedom to be absorbed, and one more abelian vector receives a
mass. Thus, if we intend to maintain an exactly massless photon in
the very end, we have to make sure that the number of gauge bosons
is at least two larger than the number of axions, which is exactly
the situation of the minimal extensions in the StSM or StMSSM, which
we introduced earlier.
\\

In the supersymmetric extension the other degrees of freedom
behave analogously. In the fermionic sector we gain a Stueckelberg
chiral fermion for each $S_j$ and a Stueckelberg gaugino for each
$C_i$. These mix with the neutral fermions of the MSSM sector
producing a neutralino mass matrix which is $4+ N_S+N_V$
dimensional. Model building with more than one Stueckelberg $U(1)$
reduces the constraints on the mixing angles and thus provides a
greater range of the parameter space for the discovery of new
physics.
\\

To illustrate this, let us briefly discuss the next simplest case,
with two extra abelian factors, and two axions. The mass matrix
for the neutral gauge fields $V^T_\mu = ( C_{\mu 3},C_{\mu 2},
B_{\mu}=C_{\mu 1}, A_{\mu}^3)$ looks then
\beqn
\phantom{ }
\hspace{.3cm}
\left[\matrix{ M_{32}^2+M_{31}^2  & M_{32}M_{22}+M_{31}M_{21} & M_{32}M_{12}+M_{31}M_{11} & 0\cr
M_{32}M_{22}+M_{31}M_{21} & M_{22}^2 +M_{21}^2 & M_{22}M_{12}+M_{21}M_{11} & 0 \cr
M_{32}M_{12}+M_{31}M_{11} & M_{22}M_{12}+M_{21}M_{11} &  M_{11}^2 + M_{12}^2 + \frac{1}{4} g_Y^2 v^2
  & - \frac{1}{4}g_Yg_2 v^2  \cr
0 &  0 & -\frac{1}{4}g_Yg_2 v^2 & \frac{1}{4}g_2^2 v^2  }\right]
\nonumber
\eeqn
where $M_{ij}$ is the \st coupling of $C_{\mu i}$ to the axion $\sigma_j$, $i=1,2,3$ and $j=1,2$, and
$C_{\mu 1}=B_\mu$. The  matrix can be written as the sum of two contributions for the \st terms and
one for the Higgs effect, each one of which has only
one non-vanishing eigenvalue, i.e. giving mass to one linear combination of gauge fields.
\\

\begin{figure}[h]
\begin{center}
\vbox{\psfig{figure=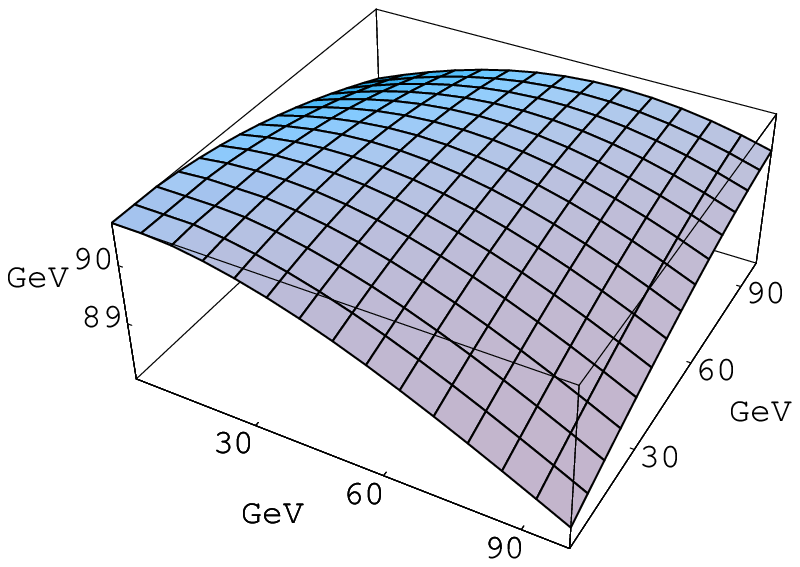,width=7cm}
\hspace{.5cm}
\psfig{figure=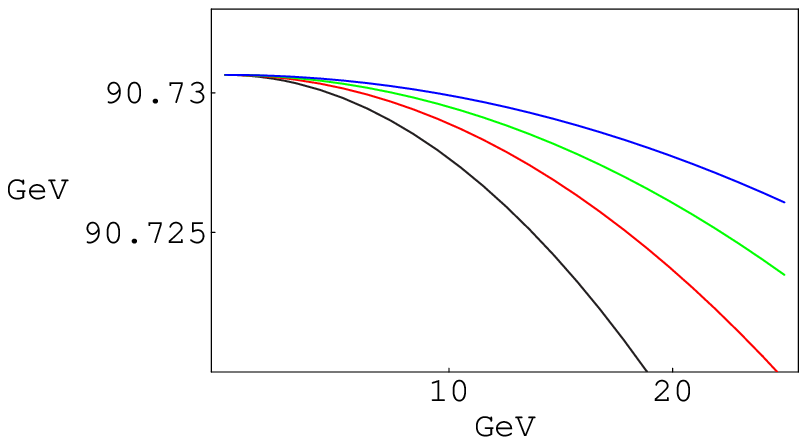,width=7cm} }
\caption{Plot of the mass of the Z, as a function of
$M_{11}$ and $M_{12}$, varying from $0$ to $100\, $GeV in the left
plot, and plotted along the variable $M_{11} = 0.6\, M_{12}$
varying from 0 to $25\, $GeV in the right one. The other mass
parameters are chosen $M_{32}=250\, \lambda\, {\rm GeV},
M_{31}=550\, \lambda\, {\rm GeV}, M_{22}=350\, \lambda\, {\rm
GeV}, M_{21}=250\, \lambda\, {\rm GeV}$. The value for $\lambda$
is 1 in the left plot, and 1 (black line), 1.3 (red line), 1.6
(green line), and 2 (blue line) in the right plot.}
\label{twou1mass}
\end{center}
\end{figure}

We can diagnonalize this matrix by an orthogonal transformation
$\co^{[1]}$ and the eigenstates $E_\mu^{[1]}=\co^{[1]} V_\mu$ are
arranged so that $E^{[1]}_\mu= ({\rm Z}_\mu'',{\rm Z}_\mu',{\rm
Z}_\mu,A^{\gamma}_\mu)^T$. The existence of two extra $U(1)$
factors now, for instance, relaxes the constraints on photonic
couplings to the hidden sector matter fields. As another
application we demonstrate that the correction to the mass of the
Z boson can now stay rather small even with comparatively large
off-diagonal terms in the mass matrix. In Figure \ref{twou1mass}
we have plotted the mass of the Z boson, the lightest
non-vanishing eigenvalue of the mass matrix above as a function of
$M_{11}$ and $M_{12}$.

The left plot shows clearly that there is a range of parameters, where the effect of turning on
the two off-diagonal elements partly cancels out, and $M_{\rm Z}$ falls off
slower than along the axes. This happens roughly along the line $M_{11} = 0.6\, M_{12}$.
In the right plot, the mass of Z is being plotted along this line, and
for various overall mass scales (measured by $\lambda$)
of the other parameters, differing by up to a factor of two.
It is evident that up to values of $25\, $GeV the effect on the Z mass is still within
some $10-30\, $MeV.
The mass eigenvalues of the matrix for $\lambda=1$ are actually
$\{ 719.4, 180.8, 90.7, 0\}$ in GeV, given the values used in Figure \ref{twou1mass}.
Together, this shows how \st extensions with multiple $U(1)$ factors have an even richer parameter space,
which involves many more options to escape experimental bounds.


\section{Conclusion}

In this paper we have given a detailed analysis of the
Stueckelberg extension of the electro-weak sector of the SM and of
the MSSM with an extra U(1) gauge group. This results in a new
heavy gauge boson ${\rm Z}'$ whose couplings to leptons and quarks
have vector and axial-vector couplings, which are different from
those for the Z boson and of the ${\rm Z}'$ bosons in conventional
$U(1)'$ extensions \cite{reviews}. Additional new features arise
for the Stueckelberg $U(1)$ extension of MSSM, where the extension
involves an abelian gauge superfield and a Stueckelberg chiral
superfield consisting. The imaginary part of the  complex scalar
is absorbed in making the $U(1)$ gauge vector massive, leaving a
spin zero scalar. It is shown that this state is heavier than the
${\rm Z}'$. Further, the neutral fermionic sector of the MSSM
extension is also significantly extended. In addition to the four
neutralino states of MSSM one has the \st chiral fermion and the
extra gaugino, which combine with the four MSSM neutral states to
produce a $6\times 6$ neutralino mass matrix. One interesting new
possibility that arises here is the case where the LSP is mostly
composed of the new fermions. In this case the lightest neutralino
of the MSSM itself will be unstable leading to a possible new
superweak candidate for dark matter. In the MSSM extension we also
considered inclusion of the Fayet-Illiopoulos D-terms and
discussed their implications.
\\

A number of phenomenological implications were discussed in section 4. It was
shown that the decay branching ratios of the ${\rm Z}'$ into quarks and
leptons are significantly different from the Z
boson, which could provide a signature for the
Stueckelberg origin of the ${\rm Z}'$. We also discussed the Higgs
sector of the extended MSSM model, where the mass matrix becomes
a $3\times 3$ matrix which mixes the residual spin zero field
of the Stueckelberg chiral
multiplet with the two CP-even neutral Higgs of MSSM. The mixings between the
MSSM Higgs and the residual Stueckelberg spin state will produce a
couplings of the latter with
visible sector fermions and its main decay mode into visible fields
is into the third generation quarks. We also discussed in section 4 the
corrections to $g_{\mu}-2$ and to sfermion masses.
\\

In section 5 we gave an analysis of some of the signatures of the
${\rm Z}'$ boson at a linear collider, such as the cross-sections
$\sigma(e^+e^-\rightarrow \mu^+\mu^-)$, $\sigma(e^+e^-\rightarrow
u\bar u)$, and $\sigma(e^+e^-\rightarrow d\bar d)$. In the
vicinity of the resonance they differ significantly from the SM
prediction. Further, the forward-backward asymmetry for the three
cases discussed above deviates sharply from the SM, again
providing an interesting signal. An interesting phenomenon is the
effect of a hidden sector on the analysis. Thus, if a hidden
sector with sufficiently light matter exists, so that the ${\rm
Z}'$ boson can decay into it, then the total width of the ${\rm
Z}'$ will be  broadened. This has drastic effect on
$\sigma(e^+e^-\rightarrow f\bar f)$, and on the forward-backward
asymmetry.
\\

In section 6 we discussed the technique for the detection of a
sharp resonance that is characteristic of the \st extension.
Finally, we have elaborated on the Stueckelberg extension of the
electro-weak sector by an arbitrary number of extra $U(1)$
factors. An interesting property of such models is the possibility
that constraints on the parameters which mix the SM gauge bosons
and the extra gauge bosons can be relaxed, allowing for the
possibility of a richer phenomenology.
\\

It should be interesting to carry out global fits to the
electro-weak data and to explore further the testability of the
\st extension  at colliders and in non-accelerator experiments.


\vspace{.5cm}
\begin{center}
{\bf Acknowledgments}
\end{center}
The authors acknowledge fruitful conversations with Tom Paul and
Darien Wood on the experimental aspects of detecting a  ${\rm
Z}'$. The work of B.~K.~was supported by the German Science
Foundation (DFG) and in part by funds provided by the U.S.
Department of Energy (D.O.E.) under cooperative research agreement
$\#$DF-FC02-94ER40818.  The work of P.~N. was supported in part by
the U.S. National Science Foundation under the grant
NSF-PHY-0139967.
\\


\end{document}